\documentclass[onecolumn,amsmath,amssymb,floatfix,10pt,aps,pre,superscriptaddress,notitlepage]{revtex4-1}

\makeatletter
\renewcommand{\p@subsection}{}
\renewcommand{\p@subsubsection}{}
\makeatother

\newcommand{\bvec}[1]{{\mathbf{\string#1} }}
\newcommand{\upd}{\mathrm{d}}
\usepackage{graphicx}
\usepackage{amssymb,amsfonts,amsmath}
\usepackage{color}
\usepackage{ulem}
\usepackage{fancybox}

\usepackage{enumitem}

\newcommand{\rr}{\boldsymbol{r}}
\newcommand{\nab}{\boldsymbol{\nabla}}

\newcommand{\xchi}{\boldsymbol{v}}

\newcommand{\taua}{\tau_\text{a}}

\newcommand{\ttau}{\tilde{\tau}}
\newcommand{\Da}{\mathcal{D}_\text{a}}

\DeclareSymbolFont{matha}{OML}{txmi}{m}{it}
\DeclareMathSymbol{\varv}{\mathord}{matha}{118}
\newcommand{\Vext}{\varv}
\newcommand{\PsiF}{f} 

\newcommand{\di}{\mathfrak{d}}

\newcommand{\Hess}{\nab\nab}

\begin{document}

\title{ Effective equilibrium states in the colored-noise model for active matter II. 
A unified framework for phase equilibria, structure and mechanical properties}
\author{Ren\'e Wittmann}
 \affiliation{Department of Physics, University of Fribourg, CH-1700 Fribourg, Switzerland}
\author{U. Marini Bettolo  Marconi}
\affiliation{Scuola di Scienze e Tecnologie, 
Universit\`a di Camerino, Via Madonna delle Carceri, 62032, Camerino, INFN Perugia, Italy}
\author{C.\ Maggi}
\affiliation{NANOTEC-CNR, Institute of Nanotechnology, Soft and Living Matter Laboratory, Piazzale A. Moro 2, I-00185, Roma, Italy }
\author{J.\ M.\ Brader} \affiliation{Department of Physics, University of Fribourg, CH-1700 Fribourg, Switzerland}

\date{\today}

\begin{abstract}
Active particles driven by colored noise can be approximately mapped onto a system that obeys detailed balance. 
The effective interactions which can be derived for such a system allow to describe the structure and phase behavior
of the active fluid by means of an effective free energy.
In this paper we explain why the related thermodynamic results for pressure and interfacial tension
do not represent the results one would measure mechanically.
We derive a dynamical density functional theory, which in the steady state simultaneously
 validates the use of effective interactions and
provides access to mechanical quantities.
Our calculations suggest that in the colored-noise model the mechanical pressure in coexisting phases 
might be unequal and the interfacial tension can become negative.
\end{abstract}

    \let\originalnewpage\newpage
    \let\newpage\relax
    \maketitle
    \let\newpage\originalnewpage

\tableofcontents

\section{Introduction}

Our understanding of equilibrium fluids tells us that all phases of a fluid at coexistence have the same pressure. 
Each phase boundary formed in the system increases its free energy, otherwise a spontaneous mixing occurs:
the interfacial tension is positive.
These familiar thermodynamic concepts need to be carefully reconsidered when the underlying time-reversal symmetry in equilibrium is broken,
e.g., by a self-propelled motion of active particles.
The most striking observation is the phase separation of colloidal particles with purely 
repulsive interactions,
only triggered by increasing their activity~\cite{buttinoni2013}.
The corresponding phase diagram is similar in form to that of an equilibrium liquid-gas transition~\cite{redner2013}, 
However, a discontinuous drop of the virial pressure at the onset of 
phase separation~\cite{winkler2015} and a negative interfacial tension~\cite{speck_interface2014}
 measured in computer simulations by the virial theorem 
underline the exceptional role of active matter.

Several different routes have been explored~\cite{takatori2015,solon_BrownianPressure2015,speckSTD,
fodor2016,solonbeyondmaxwell2016,faragebrader2015,wittmannbrader2016,marconi2015} to obtain a thermodynamic description
of the inherently non-equilibrium behavior of an active fluid in the steady state.
The phenomenology of Motility-Induced Phase Separation is now
well established in continuum models based on empirical
arguments \cite{cates2008,stenhammar2013,stenhammar2014,activeModelB,cates_tailleur2014} 
or coarse-graining strategies \cite{speck2014,speck2015}.
The driving force is a generic slow-down mechanism in the vicinity of other particles, e.g.,
due to collisions~\cite{buttinoni2013} or chemical signaling~\cite{berg1995},
resembling a passive system with attractive interactions.
Two prominent models describing these experimental observations~\cite{cates_tailleur2014} are (interacting)
active Brownian particles (ABPs) with a self-propulsion of constant magnitude in the direction of the instantaneous orientation, 
and particles propelled by a velocity-dependent swim speed.

One of the most important challenges for active thermodynamics is to understand the role of pressure.
Firstly, the notion of an active pressure is a matter of definition,
 even for a non-interacting fluid~\cite{speck2016id}.
Secondly, there is no obvious link between the thermodynamic pressure derived from a governing free energy and the
mechanical force-balance condition with the system boundaries.
 Finally, it depends on the particular model system
 whether the equality of mechanical pressure constitutes an appropriate criterion for
 phase coexistence \cite{solon_BrownianPressure2015,solonbeyondmaxwell2016}
 or whether it is a state function at all \cite{solon2015EOS}.
 Just very recently the mechanical contribution due to activity has been identified for ABPs~\cite{takatori2014,solon_BrownianPressure2015}.
 Much less is known about the chemical potential~\cite{dijkstra2016mu},
 i.e., the work necessary to insert a particle.
Understanding this quantity would be an important step to develop grand-canonical Monte-Carlo techniques required
to observe interfacial phase behavior~\cite{wittmannbrader2016}.

A model not considered in Ref.~\onlinecite{solonbeyondmaxwell2016}, but of particular theoretical appeal,
consists of particles whose self-propulsion is mimicked by a fluctuating colored-noise variable.
The resulting physics exhibit some intriguing similarities to equilibrium systems. 
For example, non-interacting particles can be described by introducing an effective temperature~\cite{szamel2014,marconi2015}
 and at low activity there exists a regime where the principle of detailed balance still holds,
 even in the presence of interactions~\cite{fodor2016}.
Going one step further, there exist convenient approximation schemes~\cite{FOX,UCNA} 
towards a system generally obeying detailed balance,
the starting point of several effective-equilibrium studies on a microscopic level
\cite{maggi2015sr,marconi2015,marconi2016mp,marconi2016sr,marconi2016,faragebrader2015,wittmannbrader2016,sharma2016,thisdraft}.
Without further empirical input it is possible to calculate the $N$-body probability distribution \cite{maggi2015sr}
and an effective interaction potential \cite{marconi2015,faragebrader2015,thisdraft} 
describing phase separation in a purely repulsive system \cite{faragebrader2015}
and related interfacial phase transition phenomena \cite{wittmannbrader2016}.
Considering a one-dimensional system at low activity, this effective-potential approximation (EPA) was shown to
perfectly agree with simulation results for the full non-equilibrium colored-noise model
in situations with~\cite{sharma2016} and without \cite{marconi2016mp} a non-vanishing probability current.
Recently, it has been demonstrated within the the effective equilibrium model that the mechanical and thermodynamical results 
for pressure and interfacial tension only coincide at lowest order in the activity \cite{marconi2016}.

In this paper we provide a new perspective on
 the EPA for the colored-noise model~\cite{thisdraft,faragebrader2015,marconi2015} 
 and demonstrate that this microscopic approach is also capable of making predictions beyond the fluid structure.
To this end we introduce the effective equilibrium approximation for the colored-noise model in Sec.~\ref{sec_CNM}
and derive in Sec.~\ref{sec_DFT} an effective dynamical density functional theory
(DDFT)~\cite{evans79,marconi1999,archerevans2004,rexloewenDDFT}
generalizing the original result of Ref.~\onlinecite{wittmannbrader2016} by including an effective diffusion tensor.
Applied in the steady state, this approach admits (i) an effective free energy~\cite{wittmannbrader2016}
yielding coexisting densities, density profiles and correlation functions,
 (ii) a mechanical stability condition which we use in Sec.~\ref{sec_PIT} to modify 
 the thermodynamic results for pressure and interfacial tension 
 to obtain a definition consistent with the measurement in simulations
and (iii) explicit calculations of these quantities without requiring further input.
 We compare in Sec.~\ref{sec_DIS} the different routes to make theoretical predictions
and discuss that our framework does, in principle, not require the crude~\cite{thisdraft,SpeckCRIT} restriction to pairwise forces.
In Sec.~\ref{sec_CON}, we conclude by comparing our findings with other theoretical frameworks.

\section{Effective equilibrium approach for the colored-noise model \label{sec_CNM}}

In the following we consider $N$ active particles ``propelled'' by
the Ornstein-Uhlenbeck processes $\xchi_i(t)$, i.e., stochastic variables
with zero mean and the non-Gaussian correlator
$ \langle\xchi_i(t)\xchi_j(t')\rangle
 \!=\!(D_\text{a}/\taua)\boldsymbol{1}\delta_{ij}\exp(-|t\!-\!t'|/\taua)$.
 The active character enters via the finite orientational decorrelation time $\taua$ and $D_\text{a}$ is the active diffusion coefficient.
The corresponding overdamped $N$-body Langevin equations read
\begin{equation}
\dot{\bvec{r}}_i(t) = \gamma^{-1}\bvec{F}_i(\bvec{r}_1,\ldots,\bvec{r}_N) 
+ \xchi_i(t)
\label{eq_OUPs}
\end{equation}
where $\gamma$ is the friction coefficient.
We assume pairwise additive interaction forces $\bvec{F}_i(\bvec{r}^N)=-\nab_i\,\mathcal{U}(\bvec{r}^N)$ 
arising from a many-body interaction potential $\mathcal{U}(\bvec{r}^N)$ 
due to one-body external fields $\Vext(\bvec{r}_i)$ and the interparticle potentials
$u(\bvec{r}_i,\bvec{r}_k)=u(|\bvec{r}_i-\bvec{r}_k|)$, such that
\begin{equation}
 \bvec{F}_i(\bvec{r}^N)= -\nab_i\bigg(\Vext(\bvec{r}_i)+\sum_{\smash{k\neq i}}u(\bvec{r}_i,\bvec{r}_k)\bigg)\,.
 \label{eq_Ftot}
\end{equation}
For the reason of simplicity, we have neglected the contribution of translational Brownian diffusion in Eq.~\eqref{eq_OUPs}.

\subsection{The multidimensional Fox approach}

Although Eq.~\eqref{eq_OUPs} does not resolve particle orientations, the non equilibrium nature of active particles 
becomes obvious in the impossibility to derive an exact Smoluchowski equation 
describing the time evolution of the probability distribution $\PsiF_N(\bvec{r}^N,t)$,
as the dynamics are always non-Markovian when a colored-noise variable $\xchi_i(t)$ is involved.
Following the multidimensional generalization~\cite{faragebrader2015,sharma2016,SpeckCRIT} of the Fox approach~\cite{FOX},
we obtain the following approximate Smoluchowski equation
\begin{align}
\frac{\partial\PsiF_N}{\partial t}=-\sum_{i=1}^N\nab_i\cdot\sum_{k=1}^N
\bvec{D}_{ik}\cdot\left(\beta\bvec{F}_k^\text{eff}-\nab_k\right)\PsiF_N
\label{eq_Smoluchowskieff}
\end{align} 
 with the inverse temperature $\beta\!=\!(k_\text{B}T)^{-1}$.
This result gives rise to
effective Markovian dynamics and thus allows an effective equilibrium description.
In Eq.~\eqref{eq_Smoluchowskieff} we identify the two central quantities of our theory.
Firstly, the effective forces
\begin{equation}
\beta\bvec{F}^{\rm eff}_k(\bvec{r}^N) = 
\sum_j\mathcal{D}^{-1}_{jk}\cdot\beta\bvec{F}_j-\nab_k\ln(\det\mathcal{D}_{[N]})\,,
\label{eq_Feff}
\end{equation}
which for an interacting system are not anymore pairwise additive, 
and, secondly, the effective diffusion tensor $\bvec{D}_{[N]}=
\mathcal{D}_{[N]}/(\beta\gamma)$ with the components
\begin{align}
\mathcal{D}^{-1}_{ij}(\bvec{r}^N)=\frac{1}{\Da}\left(\boldsymbol{1}\delta_{ij}
+\ttau\nab_i \nab_j\mathcal{U}(r^N)\right)
\label{eq_Gamma0}
\end{align}
of its inverse $\mathcal{D}_{[N]}^{-1}$, where $\ttau\!:=\!\taua/\gamma$ and $\Da\!:=\!\beta\gamma D_\text{a}$.
This quantity comprises the total contribution of activity to the system,
as it becomes trivial in the absence of activity ($\ttau=0$).
 Including the Brownian translational diffusion in Eq.~\eqref{eq_OUPs},
the form of the inverse diffusion tensor becomes more complicated~\cite{thisdraft}, compare appendix~\ref{app_Feff}.

As an alternative to the Fox approach, applying the Unified Colored Noise approximation (UCNA)~\cite{UCNA} to Eq.~\eqref{eq_OUPs}
yields another evolution equation~\cite{maggi2015sr,marconi2015,thisdraft} which differs 
from Eq.~\eqref{eq_Smoluchowskieff} by a factor $\mathcal{D}_{[N]}$.
As will become clear later, it is rather instructive that we use the Fox picture here.
In the (current-free) steady state with $\PsiF_N(r^N,t)\rightarrow P_N(r^N)$, 
however, both approximation schemes coherently yield~\cite{thisdraft}
\begin{align}
0&=\sum_j\mathcal{D}^{-1}_{jk}\cdot\beta\bvec{F}_j P_N-\nab_kP_N-P_N\nab_k\ln(\det\mathcal{D}_{[N]})\simeq\beta\bvec{F}_k^\text{eff}P_N-\nab_kP_N\,,\label{eq_ss2}
\\
0 &=\beta\bvec{F}_i P_N-\sum_k\nab_k\cdot(\mathcal{D}_{ki}P_N)\simeq\sum_k\mathcal{D}_{ki}\cdot\left(\beta\bvec{F}_k^\text{eff}P_N-\nab_kP_N\right)\,,
\label{eq_ss1}
\end{align}
where the second line is obtained after multiplying with $\mathcal{D}_{ki}$ and summing over $k$. 

\subsection{Interpretation of the two versions of the steady-state condition}
 \label{sec_CNMdiscuss}
 
Both versions, Eq.~\eqref{eq_ss2} and Eq.~\eqref{eq_ss1},  of the steady-state condition are equivalent in the sense that they result in the steady-state probability distribution~\cite{maggi2015sr} 
\begin{align}
P_N(\bvec{r}^N)\propto e^{-\beta\mathcal{H}_{[N]}(\bvec{r}^N)}\,, 
\label{eq_PN}
\end{align}
where $\mathcal{H}_{[N]}(\bvec{r}^N)$ is defined from Eq.~\eqref{eq_Feff} through
$\bvec{F}_k^\text{eff}\!=\!-\nab_k\mathcal{H}_{[N]}$.
The EPA~\cite{thisdraft,faragebrader2015,marconi2015} amounts to setting 
\begin{align}
\!\bvec{F}^\text{eff}_i\approx-\nab_i\mathcal{U}^\text{eff}=
-\nab_i\bigg(\Vext^\text{eff}(\bvec{r}_i)+\sum_{\smash{k\neq i}}u^\text{eff}(\bvec{r}_i,\bvec{r}_k)\bigg)
\label{eq_effectiveForceEPA}
\end{align}
using effective pairwise interactions
defined in appendix~\ref{app_Feff} as $u^\text{eff}(\bvec{r}_1,\bvec{r}_2)\!=\!\mathcal{H}_{[2]}(\bvec{r}_1,\bvec{r}_2)$ (assuming $\Vext\!=\!0$) and 
$\Vext^\text{eff}(\bvec{r})\!=\!\mathcal{H}_{[1]}(\bvec{r})$.

The difference in form between Eq.~\eqref{eq_ss2} and~\eqref{eq_ss1}
suggests an intriguing new interpretation of our theory.
Equation~\eqref{eq_ss2} reminds of a thermodynamic condition involving a standard ideal-gas contribution $\nab_kP_N$ and excess terms describing (activity-mediated) interactions.
We thus consider
 an effective free energy functional~\cite{marconi2015,marconi2016mp,wittmannbrader2016}
\begin{align}\label{eq_free_energy}
\mathcal{F}^\text{eff}[\,\rho\,] \!=\! \mathcal{F}_\text{id}[\,\rho\,]  +\mathcal{F}_\text{ex}^\text{eff}[\,\rho\,] 
+\int\!\mathrm{d}\bvec{r}\,\Vext^\text{eff}(\bvec{r})\,\rho(\bvec{r})\,,
\end{align} 
where the excess free energy $\mathcal{F}_\text{ex}^\text{eff}[\,\rho\,]$ follows 
from $u^\text{eff}(r)$ using standard methods (details in appendix~\ref{app_Feff}) and 
$\beta\mathcal{F}_\text{id}[\,\rho\,] \!=\! \int \mathrm{d}\bvec{r}\, \rho(\bvec{r})\left(\,\ln(\Lambda^3\rho(\bvec{r}))-1\right)$ 
is the ideal-gas term with thermal wavelength $\Lambda$ arising from $\nab_kP_N$. 
Knowing $\mathcal{F}_\text{ex}^\text{eff}$, we can define in appendix~\ref{app_Feff}
a hierarchy of direct correlation functions~\cite{evans79}
to characterize the fluid structure.
In equilibrium density functional theory (DFT), the density $\omega^\text{eff}$ of the grand potential is defined by
$\int\upd\bvec{r}\,\omega^\text{eff}(\bvec{r})\!=\!\mathcal{F}^\text{eff}[\,\rho\,] -\mu N$,
where $\mu$ denotes an effective chemical potential 
 (introduced to fix the average number of particles). 
In our approximate treatment of the active fluid we can define the 
 thermodynamic bulk pressure 
 \begin{align}
  \beta p^\text{eff}\!=\!-\omega^\text{eff}\,,
  \label{eq_peff}
 \end{align}
 where $\omega^\text{eff}$ is the grand potential density of the uniform
system and the interfacial tension 
 \begin{align}
\beta \gamma^\text{eff}\!=\!\int\upd z\left(\beta p^\text{eff}+\omega^\text{eff}(z)\right)
 \label{eq_geff}
 \end{align}
at a planar interface.
However, these formulas do not reproduce simulation results even qualitatively~\cite{wittmannbrader2016,SpeckCRIT,speck_interface2014}.

Another point of view is to consider the first equality in Eq.~\eqref{eq_ss1}
and separate external forces $\bvec{F}_i^\text{ext}\!=\!-\nab_i\Vext$ (acting on the boundary of the system)
from internal ones due to both interactions and activity.
Such a force balance allows us to identify the active pressure $p\!=\!-\mathcal{V}(\{\bvec{F}_i^\text{ext}\})/(\di V)$ by equating the 
virial~\cite{hansen_mcdonald1986}
\begin{align}
\mathcal{V}(\{\bvec{F}_i\}):=\int\upd\bvec{r}_1\ldots\int\upd\bvec{r}_N\sum_{i=1}^N\left(\bvec{F}_i\cdot\bvec{r}_i\right)P_N
\label{eq_virial}
\end{align}
of the external forces with that of the internal contributions~\cite{marconi2016}.
However, such a mechanical expression for the pressure does not necessarily allow us to study phase coexistence~\cite{solon_BrownianPressure2015,solonbeyondmaxwell2016}.

In summary, it appears that we need to consider a different condition depending on 
which property of the active fluid we are interested in.
The second expression in Eq.~\eqref{eq_ss1}, however, contains all the desired information.
Upon substituting the exact effective force~\eqref{eq_Feff}, it becomes obvious
how to properly calculate the virial of forces~\eqref{eq_virial}.
This trivial conversion will now serve to illustrate how the thermodynamical and mechanical aspects 
represented by Eqs.~\eqref{eq_ss2} and \eqref{eq_ss1}, respectively, 
can both simultaneously be cast within the framework of DDFT.

\section{Unified local force balance and free energy \label{sec_DFT}}

For the sake of generality, we now return to the dynamical problem,  Eq.~\eqref{eq_Smoluchowskieff},
derived by (i) enforcing the Markovian property.
We will further consider (ii)
 only pairwise interactions (EPA) instead of the full 
many-body effective potential, which is necessary to define the effective free energy, Eq.~\eqref{eq_free_energy},
and (iii) a diagonal diffusion tensor $\mathcal{D}_{ij}\!\approx\!\delta_{ij}\mathcal{D}^\text{p}$ with a pairwise
additive expression for $\mathcal{D}^\text{p}$ (compare appendix~\ref{app_DDFT}).
 Based on approximations (i-iii) we will now derive local force balance condition which contains as an ingredient the 
effective free energy functional, Eq.~\eqref{eq_free_energy}, derived in the EPA.

\subsection{Dynamical density functional theory for effective interactions \label{sec_DFT_R}}

Recognizing the analogy of the form of Eq.~\eqref{eq_Smoluchowskieff} with 
that of a Smoluchowski equation describing hydrodynamic interactions in a passive 
system~\cite{rexloewenDDFT,dhont},
 we integrate Eq.~\eqref{eq_Smoluchowskieff} over $N-1$ positions.
Making the adiabatic assumption of DDFT~\cite{marconi1999,archerevans2004,rexloewenDDFT} that
the correlations in the dynamic system instantaneously follow from those of an equilibrium system at given configuration,
 we derive in appendix~\ref{app_DDFT} the main result of this paper,
the following equation of motion 
\begin{equation}
\beta\gamma\frac{\partial\rho(\bvec{r},t)}{\partial t}=
\nab\cdot\boldsymbol{\mathcal{D}}(\bvec{r},t)\cdot\left(\rho(\bvec{r},t)\nab\frac{\delta\beta\mathcal{F}^\text{eff}[\,\rho\,]}{\delta\rho(\bvec{r},t)}\right)
\label{eq_DDFT}
\end{equation} 
for the time-dependent one-body density $\rho(\bvec{r},t)$, 
where $\boldsymbol{\mathcal{D}}(\bvec{r},t)$ is a dimensionless ensemble-averaged diffusion tensor.
In the following, we are interested only in the steady state and omit the time dependence.
For reasons which will become clear at the end of this section, we identify~\cite{thisdraft}
\begin{align}
\boldsymbol{\mathcal{D}}^{-1}(\bvec{r})\simeq\boldsymbol{\mathcal{D}}_\text{I}^{-1}(\bvec{r})=\frac{1}{\Da }
\left(\boldsymbol{1}+\ttau \langle\Hess\mathcal{U}\rangle\right)
\label{eq_TeffINV2}
\end{align}
with the ensemble average of Eq.~\eqref{eq_Gamma0}, where we define $\langle\mathfrak{D}\mathcal{U}\rangle:=
\mathfrak{D}\Vext(\bvec{r})+\int\upd\bvec{r}'(\rho^{(2)}(\bvec{r},\bvec{r}')/\rho(\bvec{r}))\mathfrak{D} u(\bvec{r},\bvec{r}')$ 
 for any (nontrivial) differential operator $\mathfrak{D}_i$ acting on $\bvec{r}_i$.

Our central time-evolution equation~\eqref{eq_DDFT} is equal in form to a DDFT \cite{rexloewenDDFT} for
 a passive colloidal system with hydrodynamic interactions, which
do not affect the structure and the (osmotic) pressure in equilibrium.
To check this analogy for an active system represented in terms of effective forces,
we inspect the zero-flux condition
\begin{align}
0=\boldsymbol{\mathcal{D}}(\bvec{r})\cdot\left(\nab\rho(\bvec{r})+\rho(\bvec{r})\left\langle\nab\beta \mathcal{U}^\text{eff}\right\rangle\right)\simeq\boldsymbol{\mathcal{D}}(\bvec{r})\cdot\left(\rho(\bvec{r})\nab\frac{\delta\beta\mathcal{F}^\text{eff}[\,\rho\,]}{\delta\rho(\bvec{r})}\right)\,,
\label{eq_ss3int1}
\end{align}
where the first equality is found before representing in appendix~\ref{app_DDFT} the interaction term 
with an approximate excess free energy.

Firstly, it is apparent from Eq.~\eqref{eq_ss3int1} that the fluid structure in a steady state can be characterized alone
using equilibrium DFT with effective potentials~\cite{wittmannbrader2016},
as the underlying variational principle~\cite{evans79} $\delta\mathcal{F}^\text{eff}/\delta\rho\!=\!\mu$ 
always satisfies the equality.
 Secondly, this force-balance condition suggests that the (effective) thermodynamic pressure $p^\text{eff}$ from Eq.~\eqref{eq_peff}
does \textit{not} represent the proper active pressure $p$ exerted on the system boundaries, in accordance with the discussion in Sec.~\ref{sec_CNMdiscuss}.
To illustrate this finding we consider the wall theorem~\cite{hansen_mcdonald1986}
\begin{align}
\beta p=-\int_0^\infty\upd z\,\frac{\partial\beta\Vext(z)}{\partial z}\,\rho(z)\,,
\label{eq_walltheorem}
\end{align} 
relating the bulk pressure to the density profile at a planar wall at $z\!=\!0$ with the bare potential $\Vext(z)$.
This condition is always fulfilled by the results of a (non-local) equilibrium DFT~\cite{evans79}.
In the present case, however, we employ effective potentials
and consistently obtain the pressure $p^\text{eff}$ exerted on the effective wall characterized by $\Vext^\text{eff}(z)$.
It is now obvious from Eq.~\eqref{eq_walltheorem} that $p\!\neq\!p^\text{eff}$.
The general condition in Eq.~\eqref{eq_ss3int1} consolidates the DFT approach,
as it also describes the balance of mechanical forces.
  As conjectured in Ref.~\onlinecite{wittmannbrader2016}, the averaged diffusivity $\boldsymbol{\mathcal{D}}(\bvec{r})$ 
will play a key role in the following to define the pressure from bulk properties alone.

\subsection{Equivalence of structure and pressure from different routes for an ideal gas}
 
 To further illustrate the utility of Eq.~\eqref{eq_ss3int1},
we consider non-interacting particles in an external field $\Vext(\bvec{r})$.
This problem can be solved exactly within the present model,
 as $P_N$~\eqref{eq_PN} factorizes into one-body contributions~\cite{marconi2015}.
Recognizing the equality of 
$\boldsymbol{\mathcal{D}}\!=\!\boldsymbol{\mathcal{D}}_\text{I}\!\equiv\!\mathcal{D}_{[1]}\!\equiv\!\mathcal{D}_{11}\!=\!
\Da (\boldsymbol{1}+\ttau\Hess\Vext(\bvec{r}))^{-1}
$ in this single-particle (N=1) limit~\cite{thisdraft}, we have 
 $\nab\beta\Vext^\text{eff}(\bvec{r})\!=\!\boldsymbol{\mathcal{D}}^{-1}(\bvec{r})\cdot\big(\nab\beta\Vext(\bvec{r})\!+\!\nab\cdot\boldsymbol{\mathcal{D}}(\bvec{r})\big)$
with the effective external potential $\Vext^\text{eff}$ (compare appendix~\ref{app_Feff}).
Now we plug the resulting effective free energy (with $\mathcal{F}_\text{ex}^\text{eff}\!=\!0$) into Eq.~\eqref{eq_ss3int1},
which becomes
\begin{align}
0=\boldsymbol{\mathcal{D}}\cdot\rho\nab\left(\ln\rho+\beta\Vext^\text{eff}\right)=\nab\cdot(\boldsymbol{\mathcal{D}}\rho)+\rho\nab\beta\Vext\,.
\label{eq_YBGid}
\end{align}
 The variational principle of equilibrium DFT yields the steady-state profile
$\rho(\bvec{r})\!=\!\rho_0\exp(-\beta\Vext^\text{eff}(\bvec{r}))$ in agreement with the solution of Eq.~\eqref{eq_YBGid},
 where $\rho_0\!=\!\exp(\beta\mu/\Lambda^3)$ is the bulk density.

As discussed in Ref.~\onlinecite{marconi2015} for a planar and spherical geometry 
and briefly recapitulated in appendix~\ref{app_virial},
 Eq.~\eqref{eq_YBGid} allows to identify the bulk pressure $\beta p\!=\!\Da\rho_0$
with constant bulk density $\rho_0$
or the different components of the pressure tensor at a curved surface~\cite{smallenburg2015}, 
e.g., via the external virial~\eqref{eq_virial} or the wall theorem~\eqref{eq_walltheorem}.
The effective pressure $\beta p^\text{eff}\!=\!\rho_0\!=\!\beta p/\Da$
lacks a factor $\Da$, i.e., the bulk value of $\boldsymbol{\mathcal{D}}_{zz}$, 
which, in this special case, can be absorbed into an effective temperature.
Alternatively, we can interpret the effective (ideal) pressure $p^\text{eff}$ as a local pressure~\cite{speck2016id}, 
which, in contrast to the mechanical pressure $p$, does not depend on the activity. 
 This close analogy could give a meaning to $p^\text{eff}$ also for interacting systems.

\subsection{Effective free energy from local force balance  \label{sec_DFT_V}}

In the special case discussed above, the EPA is only one possibility to derive the exact free energy and pressure,
c.f., Ref.~\onlinecite{marconi2015}. 
Moreover, Eq.~\eqref{eq_YBGid} can be seen as the first member of an active equivalent of a 
Yvon-Born-Green (YBG) hierarchy~\cite{marconi2015,marconi2016}, a mechanical force-balance equation.
In the presence of interparticle interactions such an equation can only be obtained approximately 
by considering the inverse of Eq.~\eqref{eq_Gamma0}
\begin{align}
\mathcal{D}_{ij}(\bvec{r}^N)=\Da\left(\boldsymbol{1}\delta_{ij}
-\ttau\nab_i
\nab_j\mathcal{U}(r^N)\right)
\label{eq_Gamma0invLT}
\end{align}
to leading order in $\ttau$ and integrating Eq.~\eqref{eq_ss1} over $N-1$ coordinates. The result~\cite{marconi2015,thisdraft}
\begin{align}
0=\rho(\bvec{r})\langle\nab\beta \mathcal{U}\rangle+\Da\nab\cdot\left(\rho(\bvec{r})(\boldsymbol{1}-\ttau\langle\Hess\mathcal{U}\rangle)\right)
\label{eq_YBGssA1}
\end{align}
has no obvious connection to thermodynamics:
only in the special case of Eq.~\eqref{eq_YBGid} it is possible to define a free energy, 
which allows to determine the structure (density profile) of the fluid.
In general, a free energy~\cite{marconi2015} reproducing upon functional differentiation
the (mechanical) condition in Eq.~\eqref{eq_YBGssA1} 
differs from $\mathcal{F}^\text{eff}$ identified in Eq.~\eqref{eq_ss3int1}.
Instead such a strategy amounts to a pseudo-thermodynamical picture obtained from Eq.~\eqref{eq_ss1}
by expanding $\mathcal{D}_{ki}$  with $\delta_{ki}$ being the leading order
and constructing an excess free energy including all terms but $-\nab_iP_N$.
However, there is no effective pair potential representing this artificial case, as the 
solution~\eqref{eq_PN} to Eqs.~\eqref{eq_ss2} and~\eqref{eq_ss1} is unique~\cite{thisdraft}.

To restore a unified mechanical and thermodynamic condition in the spirit of Eq.~\eqref{eq_ss3int1} from Eq.~\eqref{eq_YBGssA1}
we propose in appendix~\ref{app_Meanfield} an alternative (to the EPA) generalization of the first equality in Eq.~\eqref{eq_YBGid}
on a simple mean-field level.
The other way round, we note that a YBG-like hierarchy 
can also be obtained from integrating the first equality in Eq.~\eqref{eq_ss2} over $N-1$ coordinates~\cite{thisdraft}
in a way that is similar to the calculation in appendix~\ref{app_DDFT} (done in the more general context of DDFT).
By doing so, the averaged inverse diffusion tensor $\boldsymbol{\mathcal{D}}^{-1}_\text{I}(\bvec{r})$, which we use in Eq.~\eqref{eq_TeffINV2}, 
emerges without any approximation on the pairwise inverse tensor from Eq.~\eqref{eq_Gamma0}.
In Ref.~\onlinecite{thisdraft} this quantity is then used to establish the connection to Eq.~\eqref{eq_YBGssA1} derived from Eq.~\eqref{eq_ss1},
which motivates our choice to consider $\boldsymbol{\mathcal{D}}^{-1}(\bvec{r})\!\simeq\!\boldsymbol{\mathcal{D}}_\text{I}^{-1}(\bvec{r})$ 
in the closely related result~\eqref{eq_ss3int1} of the EPA.
An alternative motivation of this choice stems from expanding $\mathcal{D}_{[N]}(\bvec{r}^N)$ up to
linear order in $\taua$~\cite{marconi2015,marconi2016}.

Finally, we note that the presented calculations also allow us to further understand the nature of the Fox~\cite{FOX} and UCNA~\cite{UCNA} 
approaches to derive an approximate Smoluchowski equation.
Starting with the probability current of the UCNA instead of Eq.~\eqref{eq_Smoluchowskieff},
the resulting effective diffusion tensor in our DDFT equation~\eqref{eq_DDFT} would
have a completely different form~\cite{thisdraft}.
In the steady state, however, Eq.~\eqref{eq_ss3int1} equally
applies to the UCNA result, as it can be directly derived from Eq.~\eqref{eq_ss1} 
without having to introduce the adiabatic DDFT argument.
Therefore, the Fox approach presented here is more appealing for its self-consistency
with the dynamical situation.

\section{Mechanical properties of interacting active particles \label{sec_PIT}}

In Sec.~\ref{sec_DFT} we discussed that, within the EPA, 
the active pressure can be measured indirectly by calculating the density profile at a planar wall,
or, as demonstrated for an ideal gas, determined by rescaling the thermodynamic pressure
 obtained from an effective free energy $\mathcal{F}^\text{eff}$.
Our objective is to generalize the latter method for an interacting active fluid 
and relate the active pressure and interfacial tension to the thermodynamic results in Eqs.~\eqref{eq_peff} and~\eqref{eq_geff}.

\subsection{Rescaled effective pressure}

The starting point to rescale the effective mechanical properties
is the first equality in Eq.~\eqref{eq_ss3int1}
to which we apply in appendix~\ref{app_virial} the virial theorem~\cite{hansen_mcdonald1986}. 
In doing so, the effective diffusion tensor $\boldsymbol{\mathcal{D}}(\bvec{r})$ from Eq.~\eqref{eq_TeffINV2} 
allows us to recover the (averaged) external force $\bvec{F}^\text{ext}(\bvec{r})$.
Thus, we may, e.g., establish the connection to the mechanical pressure exerted on the system boundary via Eq.~\eqref{eq_virial}.
Far away from such a boundary, we argue that all internal terms which depend on the external potential vanish.
In $\di=3$ dimensions the diffusion tensor of interest for a fluid that is inhomogeneous at most in the $z$-direction 
only depends on particle interactions mediated by $u(r)$ and is diagonal.
Its components read
\begin{align}\label{eq_virialpressureDii}
&\boldsymbol{\mathcal{D}}_{\alpha\alpha}(\bvec{r})=\Da \left(1+\ttau \int\upd\bvec{r}'\,
\frac{\rho^{(2)}(\bvec{r},\bvec{r}')}{\rho(\bvec{r})}\left(\left(\partial_r^2u(r)-\frac{\partial_ru(r)}{r}\right)
\!\frac{(\bvec{r}_\alpha\!-\bvec{r}_\alpha')^2}{r^2}+\frac{\partial_ru(r)}{r}\right)
\right)^{-1},
\end{align}
 with the cartesian index $\alpha\in\{x,y,z\}$ 
 and $\partial_r\!=\!\partial/\partial r$ denotes the partial derivative with respect to $r\!=\!|\bvec{r}-\bvec{r}'|$.

\begin{figure}[t]
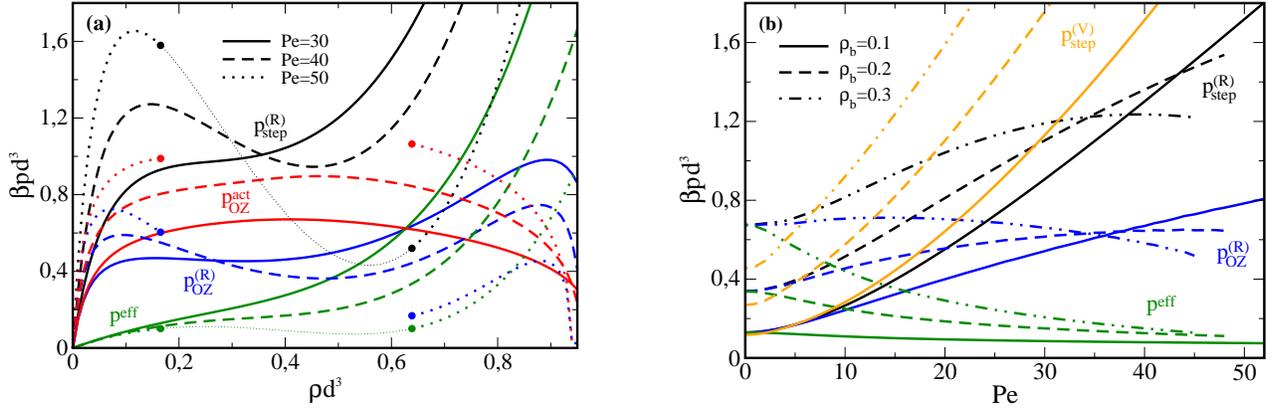

\includegraphics[height=0.3\textwidth]{Pofrho.eps}\hspace*{1.2cm}
\includegraphics[height=0.3\textwidth]{PofPeFOX.eps}
\caption{
 DFT results for the effective thermodynamic $p^\text{eff}$ and different versions of the active bulk pressure for $\beta^{-1}\ttau/d^2\!=\!0.065$.
\textbf{(a)} Dependence on the bulk density $\rho$ for different Pecl\'et numbers $Pe$.
As labeled in the subscripts, we consider our rescaled result $p^{(R)}$ from Eq.~\eqref{eq_EFFp} with the theoretical radial distribution $g_\text{OZ}(r)$ 
and the approximation $g_\text{step}(r)$ as a step function (see text), as well as  
the common first term $p^\text{act}:=\mbox{Tr}[{\boldsymbol{\mathcal{D}}}]\rho/3$ (for $g_\text{OZ}(r)$) of Eqs.~\eqref{eq_EFFp} and~\eqref{eq_EFFpYBG}.
The dots denote the coexisting densities at $Pe\!=\!50$ and the thin dotted lines denote the unstable region
(the results for $g_\text{OZ}(r)$ are unphysical in this case).
\textbf{(b)} Dependence on $Pe$ for different values of $\rho$.
 Here we show the virial pressure $p^{(V)}$ from Eq.~\eqref{eq_EFFpYBG},
calculated with $g_\text{step}(r)$,
instead of only the first term $p^\text{act}$.
At small $Pe$, the assumption to neglect the translational Brownian motion is no longer justified.
This is corrected here by adding the unit matrix to the averaged diffusion tensor in Eq.~\eqref{eq_virialpressureDii}, 
in the spirit of the Fox approach~\cite{thisdraft} (compare appendix~\ref{app_Feff}), 
such that for $Pe\!=\!0$ we have $p^{(R)}\!=\!p^\text{eff}$, i.e., the passive pressure.
}
\label{fig_p1}
\end{figure}

In a homogeneous bulk system, all components ${\boldsymbol{\mathcal{D}}}_{xx}\!=\!{\boldsymbol{\mathcal{D}}}_{yy}\!=\!{\boldsymbol{\mathcal{D}}}_{zz}\!=\!\mbox{Tr}[{\boldsymbol{\mathcal{D}}}]/3$ are equal and constant.
The rescaled active pressure becomes 
\begin{equation}
\beta p^{(R)}=\frac{\mbox{Tr}[{\boldsymbol{\mathcal{D}}}]}{3}\beta p^\text{eff}=\frac{\mbox{Tr}[{\boldsymbol{\mathcal{D}}}]}{3}\rho+\frac{\mbox{Tr}[{\boldsymbol{\mathcal{D}}}]}{3}\beta p^\text{eff}_\text{ex}\,
\label{eq_EFFp}
\end{equation}
 with the effective pressure $p^\text{eff}$ given by Eq.~\eqref{eq_peff}.
The second step illustrates the separation into ideal-gas and the excess contributions, 
where $p^\text{eff}_\text{ex}$ is identified from the virial of the effective force.
The derivation and a detailled discussion of this result can be found in appendix~\ref{app_virial}.
As it is obvious from Eq.~\eqref{eq_EFFp}, the effective contribution $\beta p^\text{eff}_\text{ex}$
to the virial pressure is different from that $\beta p^{(R)}-\rho$ in an active system.
As a consequence we can rationalize why 
the comparison of these two quantities made in Ref.~\onlinecite{SpeckCRIT} does result in a good agreement.
These simulations do not reveal a failure of the EPA in general but rather serve to justify the route taken here.

Inspecting our result~\eqref{eq_virialpressureDii} for the active pressure,
we make some interesting observations when it comes to the coexistence
of two phases at low (g) and high density (l).
In the EPA the calculation of the corresponding densities amounts to evaluating
$p_\text{(g)}^\text{eff}\!=\!p_\text{(l)}^\text{eff}$, $\mu_\text{(l)}\!=\!\mu_\text{(g)}$ and $T_\text{(l)}\!=\!T_\text{(g)}$.
Regarding Eq.~\eqref{eq_virialpressureDii} it appears that $\mbox{Tr}[{\boldsymbol{\mathcal{D}}}]$ is
a monotonously decreasing function of the density and we thus expect a higher pressure $p_\text{(g)}^{(R)}\!>\!p_\text{(l)}^{(R)}$
in the dilute phase, as it was also observed in computer simulations for active Brownian particles with periodic boundary conditions~\cite{winkler2015}.
 However, this prediction has to be taken with care, as we discuss in Sec.~\ref{sec_PITvirP}.

In order to quantify our observations, we perform some model calculations for the active pressure 
using the effective equilibrium DFT described in appendix~\ref{app_Feff}.
We fix $\beta^{-1}\ttau/d^2\!=\!0.065$ and introduce the 
 Pecl\'et number $Pe\!=\!\sqrt{3\Da\beta d^2/\ttau}$ 
as a control parameter for activity.
 Note that the mean-field approximation chosen here differs from that in Ref.~\onlinecite{wittmannbrader2016} and results in a
more realistic location of the critical point at $Pe\!\approx\! 44.531$.
In the homogeneous regime, we have $\rho^{(2)}(\bvec{r},\bvec{r}')\!=\!\rho^2g(r\!=\!|\bvec{r}-\bvec{r}'|)$,
where the radial distribution $g(r)\!=\!g_\text{OZ}(r)$ in general follows from solving the Ornstein-Zernicke equation 
 in appendix \ref{app_Feff}.
This means that for a phase-separating system the active pressure is ill-defined at intermediate densities.
We further use the simple approximation of $g_\text{step}(r)\!=\!\Theta(r/\sigma-1)$
as a step function ($\sigma$ denotes the effective hard-sphere diameter), 
which allows for an analytic calculation.

 As shown in Fig.~\ref{fig_p1}a the expected drop of the rescaled pressure at the phase transition (for $Pe\!=\!50$) is apparent for both choices of $g(r)$.
Interestingly, our model calculations suggest at intermediate activity ($Pe\!=\!40$), 
that the active pressure exhibits a loop that does not admit phase separation.
This is because the thermodynamic result $p^\text{eff}$ does not show such a loop in this case and, 
therefore, we do not find a divergence in the structure factor. 
If this was the case, the active pressure calculated with $g_\text{OZ}(r)$
would fluctuate in this region, as for $Pe\!=\!50$.
For even smaller activity, the active pressure increases monotonously with increasing density,
as expected for a non-phase-separating system.
Quite generically, we see that the increase of the active pressure at small densities becomes steeper at higher activity,
which can be expected by inspection of the leading term $\beta p^{(R)}\!\propto\!\Da \rho$ in Eq.~\eqref{eq_EFFp}.
Thus, by showing the activity-dependence of the pressure in Fig.~\ref{fig_p1}b,
we clarify that our modification of the purely thermodynamical result $p^\text{eff}$ conveniently explains the
contradiction pointed out in Ref.~\onlinecite{SpeckCRIT} with the expected behavior.
The renormalized pressure increases with increasing activity, which is now consistent with simulations.

\subsection{Comparison to the direct calculation of the virial pressure \label{sec_PITvirP}}

The strange behavior of $p^{(R)}$ at higher densities, i.e.,
the loop in the absence of phase separation and the sharp decrease when $\boldsymbol{\mathcal{D}}$ approaches zero,
as well as, the slightly non-monotonic behavior as a function of $Pe$ 
appears to be an artifact of our derivation being based on 
the simplified condition in Eq.~\eqref{eq_ss3int1} instead of that in Eq.~\eqref{eq_ss1}.
Therefore it becomes instructive to compare the rescaled result from Eq.~\eqref{eq_EFFp} to the virial pressure~\cite{marconi2016}
\begin{equation}
\beta p^{(V)}=\frac{\mbox{Tr}[{\boldsymbol{\mathcal{D}}}]}{3}\rho +\beta p_0
\label{eq_EFFpYBG}
\end{equation}
directly obtained from the first equality in Eq.~\eqref{eq_ss1} via the condition in Eq.~\eqref{eq_YBGssA1}.
The second contribution $p_0$ equals the formula for the virial pressure~\cite{hansen_mcdonald1986} in a passive system,
only depending on the bare interactions.
 It increases monotonically with increasing density $\rho$ and is independent of the activity.
In appendix~\ref{app_virial} we show how to recover this result from an empirical modification
of our more approximative approach.

The first term in Eq.~\eqref{eq_EFFpYBG} was argued~\cite{marconi2016} to be
closely related to the swim pressure~\cite{solon_BrownianPressure2015} in an active Brownian system
and is the same as that arising in Eq.~\eqref{eq_EFFp} from the effective ideal-gas pressure.
Indeed, we observe in Fig.~\ref{fig_p1}a that 
this term shows the expected 
parabolic behavior of a swim pressure~\cite{takatori2014,takatori2015,solon_BrownianPressure2015},
 This is most apparent for $g(r)\!=\!g_\text{OZ}(r)$,
 as in this case the diffusion tensor vanishes more rapidly.
Approximating once again $g(r)\!=\!g_\text{step}(r)$ as a step function, 
we obtain a simple quadratic expression $p_0\!\propto\!\rho^2$ 
for the virial pressure in Eq.~\eqref{eq_EFFpYBG}.
In contrast, the second term in Eq.~\eqref{eq_EFFp} always depends on the density in a more general form,
even when a simple radial distribution is assumed to calculate the effective diffusion tensor.

As expected, the behavior of $p^{(R)}$ and $p^{(V)}$ at small densities is observed in Fig.~\ref{fig_p1}b
to be quite similar over a large range of activities.
At very high densities, we recognize in Fig.~\ref{fig_p1}a the limitation of rescaling the effective pressure
with the effective diffusivity, since both quantities eventually vanish.
We illustrate in appendix~\ref{app_virial} that the terms resulting in the attractive part
of the effective interaction potential, which are necessary to describe the phase transition,
should cancel in the mechanical picture and thus not significantly contribute to the pressure.
This is not the case for the rescaled pressure in Eq.~\eqref{eq_EFFp}, 
as only the low-density limit of the diffusion tensor enters the effective potential.
A more detailed discussion of this point can be found in appendix~\ref{app_pressure2EPA}.

We thus conclude that the direct calculation of the virial pressure, resulting in Eq.~\eqref{eq_EFFpYBG},
is the method of choice when we are not interested in the fluid structure.
Since the contribution $p_0$ to the pressure due to bare interactions is a monotonously increasing function of the density,
 this expression is also useful for high densities.
Another striking difference is that $p_\text{(g)}^{(V)}\!<\!p_\text{(l)}^{(V)}$ appears to jump to a higher value at coexistence,
 which we already observe in Fig.~\ref{fig_p1}a for the common first term of $p^{(R)}$ and $p^{(V)}$.
On the downside, as we require the radial distribution function as an input to calculate $p^{(V)}$,
the results from this approach are not self-consistent.
It is thus not possible to make a definite statement of how active pressure behaves in our model
without going beyond present approximations.
 Another scenario is that of an equal pressure at coexistence, which is suggested by calculating the pressure on a planar wall following Eq.~\eqref{eq_walltheorem}, 
when taking into account the wetting transition predicted in Ref.~\onlinecite{wittmannbrader2016}.
This behavior agrees with the expectation for ABPs~\cite{solon_BrownianPressure2015,solonbeyondmaxwell2016}.

\subsection{Rescaled and virial formulas for the interfacial tension}

Considering now a planar geometry, we define 
$\boldsymbol{\mathcal{D}}_{\alpha\alpha}(z)\!:=\!\iint\upd x\upd y\,\boldsymbol{\mathcal{D}}_{\alpha\alpha}(\bvec{r})$ 
with $\boldsymbol{\mathcal{D}}_{xx}\!=\!\boldsymbol{\mathcal{D}}_{yy}\!\neq\!\boldsymbol{\mathcal{D}}_{zz}$
and derive in appendix~\ref{app_virial} the rescaled active interfacial tension 
\begin{align}
\label{eq_EFFgamma}
\beta \gamma^{(R)}=\int\upd z&\left(\boldsymbol{\mathcal{D}}_{zz}(z)\beta p^\text{eff}
+\boldsymbol{\mathcal{D}}_{xx}(z) \omega^\text{eff}(z)+\rho(z)\,z\,\partial_z\boldsymbol{\mathcal{D}}_{zz}(z)\right)
\end{align}
at the free interface, where we identify the negative grand potential density $\omega^\text{eff}(z)$ 
with the effective tangential pressure.
With unequal bulk pressures at phase coexistence, it is obvious that in our model also the normal pressure, 
i.e., the terms depending on $\boldsymbol{\mathcal{D}}_{zz}$, is not a constant along the interface.
However, as expected, there is no contribution to the active interfacial tension $\gamma^{(R)}$ in Eq.~\eqref{eq_EFFgamma} 
at $|z|\gg0$ sufficiently far away from the interface, 
as in the bulk all components of the effective diffusion tensor are equal and constant.

Again, it is instructive to compare Eq.~\eqref{eq_EFFgamma} to the result~\cite{marconi2016}
\begin{equation}
\beta \gamma^{(V)}=\int\upd z\left(\boldsymbol{\mathcal{D}}_{zz}(z)-\boldsymbol{\mathcal{D}}_{xx}(z)\right)\rho(z)
+\beta \gamma_0 
\label{eq_EFFgammaYBG}
\end{equation}
derived from Eq.~\eqref{eq_YBGssA1},
where $\gamma_0\!>\!0$ is a standard virial expression
 by Kirkwood and Buff~\cite{kirkwoodbuff,marconi2016} (see appendix~\ref{app_virial}), 
 which is independent of the activity.
As discussed in appendix~\ref{app_virial},
the first two terms in Eq.~\eqref{eq_EFFgamma} imply a similar separation into 
an ideal-gas-like term 
and an effective virial contribution, which after switching to the DFT picture is not explicit any more.
As a major difference to Eq.~\eqref{eq_EFFgammaYBG}, we identify in the last term in Eq.~\eqref{eq_EFFgamma} an additional contribution to the
normal pressure, which vanishes in either bulk state.
The presence of this term is explained in appendix~\ref{app_virial}
to counteract a similar contribution arising from the effective interaction potential
in an approximate way.

\begin{figure}[t]{
\includegraphics[height=0.172\textwidth] {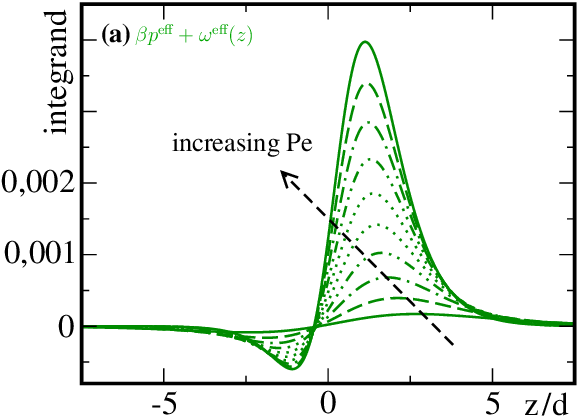} \hfill
\includegraphics[height=0.172\textwidth] {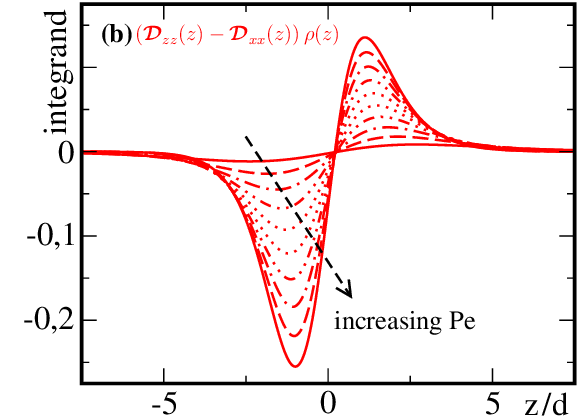} \hfill
\includegraphics[height=0.172\textwidth] {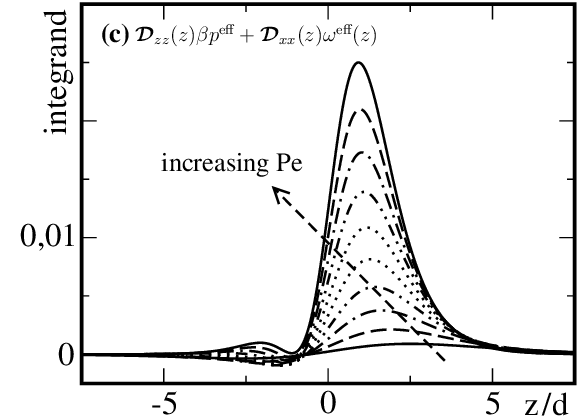} \hfill
\includegraphics[height=0.172\textwidth] {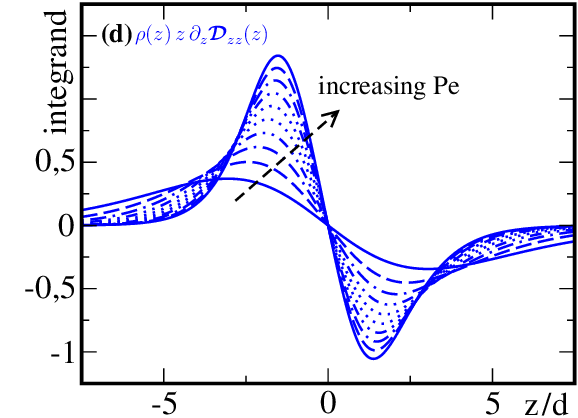}}

\caption{ Different contributions to the integrand of the interfacial tension calculated with $g_\text{step}(r)$
for different Pecl\'et numbers increased from $Pe\!=\!46$ to $Pe\!=\!55$ in steps of 1.
\textbf{(a)} Effective thermodynamic integrand of $\gamma^\text{eff}$ without rescaling.
\textbf{(b)} Ideal contribution to $\gamma^{(V)}$, Eq.~\eqref{eq_EFFgammaYBG}.
\textbf{(c)} Rescaled thermodynamic integrand in Eq.~\eqref{eq_EFFgamma}.
\textbf{(d)} Correction for $\gamma^{(R)}$ by the last term in Eq.~\eqref{eq_EFFgamma}.
\label{fig_g}}
\end{figure}

\begin{table*}[t]
\centering {\large Markovian ``effective equilibrium'' approximation 
for the {colored-noise model~\eqref{eq_OUPs}}}
\begin{tabular}{ccccc}
&     &  $\large\pmb{\downarrow}$    &   \parbox{0.03cm}{\vspace*{0.5cm}}  &  \\ \vspace*{-0.1cm}  
  \Ovalbox{\Ovalbox{\parbox{4.85cm}{ \textbf{mechanical picture of the steady-state condition~\eqref{eq_ss2}}}}}  & \parbox{0.91cm}{$\pmb{\Longleftarrow}$ \\\vspace*{0.2cm} $\pmb{\Longleftrightarrow}$}   & \parbox{4cm}{ \textbf{effective dynamics~\eqref{eq_Smoluchowskieff}} \\\vspace*{0.2cm} $\mathcal{D}_{[N]}(\bvec{r}^N)$~\eqref{eq_Gamma0}}   &   \parbox{0.91cm}{\vspace*{0.3cm}$\pmb{\Longrightarrow}$ \\\vspace*{0.2cm} $\pmb{\Longleftrightarrow}$\vspace*{0.3cm}}   &  \Ovalbox{\Ovalbox{\parbox{4.85cm}{\textbf{thermodynamical picture of the steady-state condition~\eqref{eq_ss1}}}}}\\
    $\!\vdots$    &     &     &     & $\!\vdots$  \\ \vspace*{-0.05cm} 
 low-activity limit~\cite{marconi2015,marconi2016,thisdraft}  &     &  \parbox{4.6cm}{\vspace*{0.4cm}}   &     &   one/two-particle limit~\cite{faragebrader2015,wittmannbrader2016,thisdraft} \\ 
  $\pmb{\downarrow}$  &     &   \parbox{4.6cm}{\vspace*{0.5cm}}  &     &  $\pmb{\downarrow}$ \\ \vspace*{-0.12cm} 
  \framebox{\parbox{4.85cm}{pairwise diffusion tensor~\eqref{eq_Gamma0invLT}}} &   $\ldots$  & \parbox{4.29cm}{\textit{minimal requirement to construct a closed theory}}   &  $\ldots$   & \framebox{\parbox{4.85cm}{pairwise effective force~\eqref{eq_effectiveForceEPA}}}\\ \vspace*{-0.05cm}  
   $\!\vdots$  &     &     &     & $\!\vdots$ \\ \vspace*{-0.1cm} 
       YGB-like hierarchy~\cite{marconi2015,marconi2016,hansen_mcdonald1986}  &     &     &     &   DFT implementation of EPA~\cite{evans79,rosenfeld89} \\ \vspace*{-0.05cm} 
  $\pmb{\Downarrow}$    &     &   \parbox{4.6cm}{\vspace*{0.6cm}}  &     & $\pmb{\Downarrow}$\\ \vspace*{-0.075cm} 
  \doublebox{\parbox{4.85cm}{  local force-balance~\eqref{eq_YBGssA1}\begin{itemize}[leftmargin=.6cm]\vspace*{-0.025cm}                                                                                                                                                                                                                                           \item[$\pmb{+}$]pressure~\eqref{eq_EFFpYBG}\vspace*{-0.06cm} \item[$\pmb{+}$]interfacial tension~\eqref{eq_EFFgammaYBG}\end{itemize}}}  & $\pmb{\longleftarrow}$    &   \parbox{4.6cm}{\textit{ EPA provides 
  input for mechanical formulas}~(\ref{eq_EFFp}-\ref{eq_EFFgammaYBG}) \textit{ and the wall theorem~\eqref{eq_walltheorem}}}  &  $\pmb{\Longleftarrow}$   & 
  \doublebox{\parbox{4.85cm}{effective free energy (appendix~\ref{app_Feff}) \begin{itemize}[leftmargin=.6cm]  \vspace*{-0.025cm}                                                                                                                                                                                                                                         \item[$\pmb{+}$] structure~\cite{faragebrader2015,wittmannbrader2016,marconi2016mp}\vspace*{-0.06cm}  \item[$\pmb{+}$] phase behavior~\cite{faragebrader2015,wittmannbrader2016}                                                                                                                                                                                                                                                                                                   \end{itemize}}}
  \\ \vspace*{-0.075cm} 
    $\!\vdots$  &     &     &     & $\!\vdots$ \\ \vspace*{-0.05cm} 
   $\times\,\boldsymbol{\mathcal{D}}(\bvec{r})\cdot\boldsymbol{\mathcal{D}}^{-1}(\bvec{r})$  &     &   \parbox{4.6cm}{\vspace*{0.6cm}}  &     & pairwise $\mathcal{D}_{ij}(\bvec{r}^N)\propto\delta_{ij}$~\eqref{eq_appDij}  \\ \vspace*{-0.1125cm} 
   $\pmb{\Downarrow}$   &     &     &     &  $\pmb{\downarrow}$  \\ \vspace*{-0.075cm} 
    mean-field approximation   &   $\pmb{\longrightarrow}$   &  \parbox{4.6cm}{\textbf{approximate unified local force balance~\eqref{eq_ss3int1}/\eqref{eq_YBGssA2}}}   &   $\pmb{\Longleftarrow}$     &  DDFT~\eqref{eq_DDFT} with $\boldsymbol{\mathcal{D}}(\bvec{r})$ \\ \vspace*{-0.4cm} 
   $\pmb{\downarrow}$   &     &    &   & $\pmb{\Downarrow}$  \\ 
       &     &   $\!\vdots$  &    &   \\\vspace*{-1.1cm}   &     &     &     &\\ 
 \framebox{\parbox{4.85cm}{effective free energy (appendix~\ref{app_Meanfield})\begin{itemize}[leftmargin=.6cm]\vspace*{-0.025cm}\item[$+$]not linear in activity\vspace*{-0.06cm} \item[$-$]only known implicitly\vspace*{-0.06cm} \item[$-$] mean-field structure \end{itemize}}}     &  \parbox{0.91cm}{\vspace*{0.3cm}$\ldots$ \\\vspace*{0.65cm} \mbox{}\vspace*{0.75cm}}   & \parbox{4.6cm}{\vspace*{0.75cm}\textit{two routes are inconsistent}\\\vspace*{0.04cm}\pmb{$\downarrow$}\\\vspace*{0.1125cm}\framebox{\parbox{4.4cm}{low-activity limit for EPA~\cite{thisdraft}\begin{itemize}[leftmargin=.6cm] \vspace*{-0.025cm} \item[+]better virial term (app.~\ref{app_pressure2EPA})\vspace*{-0.06cm}  \item[$-$]improper thermodynamics \end{itemize}}}} 
  &  \parbox{0.91cm}{\vspace*{0.3cm}$\ldots$ \\\vspace*{0.65cm} $\pmb{\longleftarrow}$\vspace*{0.75cm}}  &  \framebox{\parbox{4.85cm}{local force-balance~\eqref{eq_ss3int1}\begin{itemize}[leftmargin=.6cm] \vspace*{-0.025cm} \item[+]pressure~\eqref{eq_EFFp}\vspace*{-0.06cm} \item[+]interfacial tension~\eqref{eq_EFFgamma}\vspace*{-0.06cm} \item[$-$]wrong virial contribution \end{itemize}}} 

 \end{tabular}
  \caption{Overview of different approaches to make explicit calculations 
  within the effective equilibrium approximation for the colored-noise model,
  i.e., Eq.~\eqref{eq_Smoluchowskieff}.
  The arrows with double line denote exact operations (or established methods)
  and those with a single line denote approximate operations.
  Each approximation is also necessary for the subsequent steps.}
 \label{tab_ILLUSTRATION1}
 \end{table*}

In Fig.~\ref{fig_g} we show how the integrand of the interfacial tension is modified by increasing activity.
As for the pressure, we employ the DFT from appendix~\ref{app_Feff} with the averaged diffusion tensor in Eq.~\eqref{eq_virialpressureDii}
and use $g_\text{step}(r)\!=\!\Theta(r/\sigma-1)$.
In contrast to the effective thermodynamic interfacial tension,
with the integrand shown in Fig.~\ref{fig_g}a,
the contribution of the activity-modified ideal-gas term in Eq.~\eqref{eq_EFFgammaYBG} is highly negative 
and outweighs the smaller positive virial contribution $\gamma_0$ (not shown).
Therefore, the overall interfacial tension $\gamma_\text{YBG}$ becomes negative~\cite{speck_interface2014}
and Fig.~\ref{fig_g}b suggests a decrease with increasing activity.
However, the behavior of this modified ideal-gas contribution does not become manifest implicitly in Eq.~\eqref{eq_EFFgamma}.
In contrast, both contributions to the integrand of $\gamma$, shown in Fig.~\ref{fig_g}c and Fig.~\ref{fig_g}d,
are positive and increase with increasing activity.

\section{Summary \label{sec_DIS}}

We developed in Sec.~\ref{sec_CNMdiscuss} a new interpretation of the steady-state condition in the effective equilibrium 
approximation for the colored-noise model.
With the goal to arrive at a predictive theory, it becomes necessary to integrate the steady-state conditions 
and thereby consider approximate pairwise additive quantities.
There are different ways to do so, as reviewed in the first paper of this series~\cite{thisdraft} and resumed in Sec.~\ref{sec_DFT}
with the focus on mechanical properties of the active fluid.
The degree of accuracy, discussed in Sec.~\ref{sec_PIT}, of the thermodynamic route (``rescaled'' formulas based on Sec.~\ref{sec_DFT_R})
and the mechanical route (``virial'' formulas based on Sec.~\ref{sec_DFT_V})
is reversed when it comes to characterize the structure of the fluid.
The interconnection of these two routes is explained in the following and illustrated in Table~\ref{tab_ILLUSTRATION1}.

\subsection{Connection between mechanic and thermodynamic route}

The form of Eq.~\eqref{eq_ss2} corresponds to a thermodynamic condition 
to determine the structure and phase behavior in the steady state.
This enables the interpretation of an effective attraction as the driving force of phase separation, 
which on the two-particle level can be cast in an effective pair potential~\cite{faragebrader2015,thisdraft}.
However, even if it was possible to represent the exact effective interaction force within a thermodynamic free energy,
we explained that the thermodynamic results for pressure and interfacial tension cannot coincide with the 
respective definitions based on the balance of mechanical forces.
This reflects the non-equilibrium nature of active systems, even in our effective equilibrium model,
and is not a consequence of assuming only a pairwise effective force as argued in Ref.~\onlinecite{SpeckCRIT}.

The proper way to switch to the mechanical picture provided by Eq.~\eqref{eq_ss1}
in order to calculate pressure and interfacial tension 
is to multiply Eq.~\eqref{eq_ss2} with the effective diffusion tensor,
ensuring that the bare interaction forces are recovered from the effective ones
that are responsible for the fluid structure.
This marks the fundamental difference to a passive system with hydrodynamic interactions (described by an evolution equation similar to Eq.~\eqref{eq_Smoluchowskieff}),
where thermodynamical and mechanical pressure are equal and the diffusion tensor does not contribute to the pressure.
We argued that the effective ``thermodynamic'' attraction 
originating from the term $\nab_k\ln(\det\mathcal{D}_{[N]})$ 
does not result in an equal decrease of the mechanical pressure.
We stress that these theoretical insights 
are in general not restricted to the (highly criticized~\cite{SpeckCRIT}) approximations involved in defining pairwise effective forces~\cite{thisdraft}
to simplify the general condition in Eq.~\eqref{eq_ss1}.

\subsection{Inconsistency arising from pairwise approximations}

Employing a pairwise approximation eventually results in an inconsistency between the thermodynamic route based on Eq.~\eqref{eq_ss2} 
and the mechanical route based on Eq.~\eqref{eq_ss1}, which becomes manifest in either direction.
On the one hand, the approach based on the EPA (first introducing an effective free energy to be later recovered in an approximate local force balance equation from Eq.~\eqref{eq_ss3int1})
seems to underestimate the virial pressure in Eq.~\eqref{eq_EFFp}.
A perhaps more accurate formula~\eqref{eq_EFFpYBG} can be derived directly from
a YBG-like force-balance equation in Eq.~\eqref{eq_YBGssA1}.
On the other hand, as a prerequisite to do explicit calculations,
we require $g(r)$ and therefore an implicit knowledge of the phase diagram,
which can only be provided by an appropriate free energy.
In this sense, the EPA is particularly appealing as it allows both defining Eq.~\eqref{eq_EFFp} 
(or Eq.~\eqref{eq_EFFpYBG}) and providing the required input.
In appendix~\ref{app_pressure2EPA} we discuss how a different choice of the effective diffusion tensor 
could improve the mechanical EPA results.
 Vice versa, we derive in appendix~\ref{app_Meanfield} a free energy by combining Eq.~\eqref{eq_YBGssA1} with the effective diffusion tensor from Eq.~\eqref{eq_virialpressureDii},
such that Eqs.~\eqref{eq_EFFp} and~\eqref{eq_EFFpYBG} become practically equivalent.
However, due to its mean-field nature, such a free energy cannot be used to predict the fluid structure, 
in contrast the one derived from the EPA.

Another interesting difference between the mechanical and thermodynamical picture
lies in the validity of low-activity expansions.
In the thermodynamic picture, the expansion in the parameter $\ttau$ is not justified~\cite{thisdraft} because of
the finite radius of convergence of the logarithmic term in Eq.~\eqref{eq_ss2}.
However, it was shown~\cite{marconi2016,thisdraft} that at first order of such an expansion 
the mechanical and the thermodynamical (without the EPA) route to define pressure and interfacial tension are consistent.
Moreover, for the EPA we discuss in appendix~\ref{app_pressure2EPA} that at this level 
the spurious negative contribution in the formula for the effective virial pressure, 
which arises from the logarithmic term in the effective force, 
can be compensated.
As the expansion in $\ttau$ converges for all terms present in Eq.~\eqref{eq_ss1}, we understand that 
 such an approximation is suitable to derive formulas for mechanical quantities.
Hence, we see that it is a difficult task in future work to arrive at a higher level of self-consistency
between the typical approximations to simplify Eq.~\eqref{eq_ss1} or Eq.~\eqref{eq_ss2},
which cannot be achieved by including higher-order terms in $\ttau$
(see the note in Table~\ref{tab_ILLUSTRATION1}).

\subsection{Outlook}

Despite its simplicity, our DDFT in Eq.~\eqref{eq_DDFT} clearly demonstrates
the importance of switching between a mechanical and thermodynamical
interpretation of the steady-state condition obtained for our model.
 We stressed that the perhaps most important advantage of the thermodynamic route is the possibility to determine the fluid structure as input for the results obtained from the mechanical route.
 As an alternative to the resulting bulk formulas based on the radial distribution, the bulk pressure
can be calculated from the profile of the one-body density at a wall, cf., Eq.~\eqref{eq_walltheorem}, which can be easily calculated with the help of the EPA. 
It is also interesting to investigate the force on a curved~\cite{smallenburg2015} or structured~\cite{nikola2016} wall along these lines.

Our results should further guide the way to perform more elaborate numerical calculations.
For example, the ``exact'' simulation of passive particles interacting with the effective potential
could answer some open questions regarding the behavior of the pressure at coexistence.
Moreover, one can determine pair~\cite{pairint} (and higher-order~\cite{tripletint}) interaction potentials 
(which may depend on the density) from a many-body simulation of any suitable active system, 
or simply take the structural information to calculate the pressure and interfacial tension
with the formulas derived from a YBG-like condition in Ref.~\onlinecite{marconi2016} and recovered here.
With such a numerical insight, we could also improve the purely theoretical approach 
by incorporating some semi-empirical corrections of the free energy
to account for (many-body) effects ignored in the definition of the effective pair potential.

\section{Conclusions \label{sec_CON}}

In this paper we have explicitly shown that
there is no equal thermodynamical and mechanical interpretation 
of the effective steady-state condition derived for active particles propelled by colored noise.
This main conclusion agrees with the statement of Ref.~\onlinecite{solonbeyondmaxwell2016},
established on a more coarse-grained level for an arbitrary model system.
For the study of mechanical quantities of the (non-equilibrium) steady state, the appropriate starting point is Eq.~\eqref{eq_ss1},
as the bare interaction forces are separated from all activity-dependent quantities.
On the other hand, a thermodynamic theory requires that there is an ordinary 
passive ideal-gas contribution as in Eq.~\eqref{eq_ss2},
as the "reference term" to construct a free energy.
 In general, we established and quantified that
the effective thermodynamical pressure is unequal, although related,
to the pressure that an active fluid exerts on its boundary.

Within the EPA, we approximately separated the governing equation~\eqref{eq_ss3int1} of the steady state
including a purely thermodynamic condition and a mechanical ``correction'' factor $\boldsymbol{\mathcal{D}}(\bvec{r})$.
This average diffusion tensor exhibits certain similarities with the concept of an effective temperature.
However, as a result of its position dependence and tensorial nature, 
it is not constant over the whole system
but remains irrelevant for calculating phase equilibria. 
According to the interpretation chosen here 
all Boltzmann factors just comprise the ordinary thermal weight $\beta$, as in Eq.~\eqref{eq_PN},
whereas all activity dependence relevant for the fluid structure is captured by effective interactions~\cite{thisdraft}.
Only the effective thermodynamic results for pressure and interfacial tension 
are then to be rescaled according to \eqref{eq_EFFp} or \eqref{eq_EFFgamma}.
In this step it does not appear to be sufficient to use the constant factor $\Da$, 
which is commonly used to define the effective
temperature~\cite{szamel2014,maggi2015sr,marconi2015,marconi2016mp,marconi2016sr,marconi2016}
and equals $\beta^{-1}\boldsymbol{\mathcal{D}}_{\alpha\alpha}(\bvec{r})$ only in the absence of any interaction
(or in the special case of a linear external potential~\cite{marconi2015}).

Our generalized DDFT~\eqref{eq_DDFT} admits a
similar form of the one-body current as found phenomenologically by 
taking into account a density-dependent swim speed~\cite{stenhammar2013,stenhammar2014}. 
This approach does, however, not admit
a direct connection to a force-balance equation in the steady state, as it has been established here.
Moreover, in our derivation we obtained from first principles a tensorial diffusivity and a
non-local free energy~\cite{wittmannbrader2016} without any empirical input.
It would be interesting to apply our approach to study, for example, the coarsening dynamics 
in the early state of phase separation.
In the steady state this analogy suggests that the approximate connection between the excess 
chemical potential and the swim pressure found in Ref.~\onlinecite{wittmannbrader2016} is 
consistent but requires further investigation.

Since the effective equilibrium system constructed from the EPA respects detailed balance,
we can further make contact to the discussion in Ref.~\onlinecite{activeModelB}.
The phase separation is driven by (effective) attractive interactions, whereas we showed that
the effective diffusivity does not affect phase coexistence.
This quantity thus plays a role similar to that of non-integrable gradient terms in the probability current. 
Coherent with our (approximate) observation from the EPA, the equivalence of thermodynamic pressure in such a situation close to equilibrium
can be argued~\cite{activeModelB} to result in a discontinuous mechanical pressure. 
Despite this observation our Eq.~\eqref{eq_EFFp} suggests that the rescaled pressure is a 
(wall-independent) state function.
In order to clarify this seemingly contradiction with some findings for 
different model systems~\cite{solonbeyondmaxwell2016,solon2015EOS}
and identify the role of a swim pressure~\cite{solon_BrownianPressure2015,takatori2014}
and active chemical potential~\cite{dijkstra2016mu} in the EPA,
it would be enlightening to integrate active particles propelled by colored noise, 
cf., Eq.~\eqref{eq_DDFT}, into a generalized thermodynamical framework.

\section*{Acknowledgements}

R.\ Wittmann and J.\ M.\ Brader acknowledge funding provided by the Swiss National Science Foundation.
C.\ Maggi acknowledges support from the European Research Council under the European Union’s Seventh Framework programme
(FP7/2007-2013)/ERC Grant agreement No. 307940.
We acknowledge A.\ Sharma for stimulating discussions.
\newline

\appendix

\section{Effective free energy for the full Fox approach \label{app_Feff}}

In the main text, we derived the probability distribution $P_N(\rr^N)$ \eqref{eq_PN}.
In the single-particle case $N\!=\!1$, we can thus identify from $P_1(\rr)\!=\!\exp(-\beta\Vext^\text{eff}(\rr))$ the effective external field
\begin{align}
\nab\beta \Vext^\text{eff}(\rr)=\mathcal{D}^{-1}_{[1]}\cdot\left(\nab\beta\Vext(\rr)+\nab\cdot\mathcal{D}_{[1]}\right)\,,
\label{eq_Veff}
\end{align}
where $\Vext(\rr)$ denotes the bare interaction of a passive particle.
  For curved potentials~\cite{nikola2016,smallenburg2015}, the second term can be conveniently replaced by different approximate forms.
This makes the EPA, i.e., the expression on the left-hand side of Eq.~\eqref{eq_YBGid} of the main text more suitable for such situations.

Considering a system of $N\!=\!2$ particles interacting with the bare potential $u(r)$ and neglecting the external forces, 
we find the effective pair potential 
\begin{align}
\nab_i\beta u^\text{eff}(r)
=\mathcal{D}^{-1}_{ii}\cdot\left(\nab_i\beta u(r)+\nab_i\cdot\mathcal{D}_{ii}\right)
\label{eq_ueff}
\end{align}
in the diagonal approximation~\cite{thisdraft}.
Another way to derive these effective potentials is to solve Eq.~\eqref{eq_ss2} 
for the respective number of particles~\cite{faragebrader2015,thisdraft}.
This strategy becomes necessary if we do not neglect the Brownian translational diffusivity in Eq.~\eqref{eq_OUPs} 
and make the Fox approximation, which then does not admit a closed-form result for $P_N$.

In order to be consistent with our earlier calculations~\cite{faragebrader2015,wittmannbrader2016},
we will construct an excess free energy functional $\mathcal{F}_{\rm ex}^{\rm eff}[\,\rho\,] $ 
from the effective potential $u^\text{eff}(r)$,
where we approximate the effective diffusion tensor in the Laplacian form~\cite{thisdraft} as
\begin{align}
\mathcal{D}_{ij}\approx\delta_{ij}\boldsymbol{1}\left(1+\Da\,\frac{r^2}{r^2+\ttau \partial_r(r^2 \partial_r u(r))}\right).
\end{align}
The first term stems from the translational diffusion
and the second term constitutes an approximation for Eq.~\eqref{eq_Gamma0} of the main text.
The effective potential is then separated into a repulsive and an attractive contribution~\cite{WCA,wittmannbrader2016}.
The first gives rise to an effective hard-sphere diameter \cite{barker_henderson1967,wittmannbrader2016}
\begin{align}
 \sigma=\int_0^{r_0} \mathrm{d}r\left(1-e^{-\beta u^{\rm eff}(r)+\beta u^{\rm eff}(r_0)}\right)
\end{align}
where $r_0$ is the position of the minimum of $u^{\rm eff}(r)$.
This term will be treated as the excess free energy
$\mathcal{F}_{\rm ex}^{\rm (hs)}[\,\tilde{\rho}\,]$ of hard spheres 
at the rescaled density $\tilde{\rho}\!=\!\rho\, d^3\!/\sigma^3$,
where we employ Rosenfeld's fundamental measure theory \cite{rosenfeld89}.

Finally, we obtain 
\begin{align}
\!\!\!\!\!\mathcal{F}_{\rm ex}^{\rm eff}[\,\rho\,] = 
\mathcal{F}_{\rm ex}^{\rm (hs)}[\,\tilde{\rho}\,] 
+\! {\iint}\! \mathrm{d}\rr_1 \mathrm{d}\rr_2 \rho(\rr_1)\rho(\rr_2){\frac{u^{\rm eff}_{\rm att}(r_{12})}{2}}  \label{meanfield}
\end{align}
 in a generalized mean-field approximation, where $r_{12}\!=\!|\rr_1-\rr_2|$ and 
 $u_{\rm att}^{\rm eff}(r)\!=\!0$ if $r\!<\!\sigma$,
 $u_{\rm att}^{\rm eff}(r)\!=\!u^{\rm eff}(r_0)$ if $\sigma\!<\!r\!<\!r_0$ and
$u_{\rm att}^{\rm eff}(r)\!=\!u^{\rm eff}(r)$ otherwise. 
By making this choice for the attractive part $u_{\rm att}^{\rm eff}(r)$ of the effective potential,
we implicitly assume the form of a step function for the radial distribution function of the reference fluid
instead of setting it to unity as in Ref.~\onlinecite{wittmannbrader2016}.
Given any excess free energy functional, we calculate the direct correlation functions~\cite{evans79}
\begin{align}
 c^{(n)}(\bvec{r}_1,\ldots,\bvec{r}_n)=-\frac{\delta^n\beta\mathcal{F}_\text{ex}^\text{eff}[\,\rho\,]}{\delta\rho(\bvec{r}_1)\ldots\delta\rho(\bvec{r}_n)}
\end{align}
of order $n$.
When we are interested in the pair structure ($n\!=\!2$),
we find the radial distribution function $g(r_{12})$ of the homogeneous fluid
from an iterative solution of the Ornstein-Zernicke equation~\cite{hansen_mcdonald1986}
\begin{align}
\!\!\!\!g(r_{12})=1+c^{(2)}(r_{12})\, \rho \int\upd\rr_3\,c^{(2)}(r_{13}) (g(r_{23})-1)\,. 
\end{align}
This result is required to calculate the averaged diffusion tensor in Eq.~\eqref{eq_virialpressureDii} of the main text, 
and, therefore, the active pressure and interfacial tension.

\section{Effective dynamical density functional theory \label{app_DDFT}}

In this appendix we demonstrate how the approximate DDFT in Eq.~\eqref{eq_DDFT}
can be deduced from Eq.~\eqref{eq_Smoluchowskieff}.
As stated in the main text this starting point represents an approximate Markovian time evolution
of active particles driven by Gaussian colored noise.
We further assume 
\begin{align}
\mathcal{D}_{ij}(\rr^N)\approx\delta_{ij}\sum_{l\neq i}^N\mathcal{D}^\text{p}(\rr_i-\rr_l)
\label{eq_appDij}
\end{align}
with the yet unspecified contributions $\mathcal{D}^\text{p}$ used to construct an
approximate pairwise additive representation for the diagonal components $\mathcal{D}_{ii}$ of the effective diffusion tensor.
Note that $\mathcal{D}^\text{p}$ should also contain one-body terms,
which can be treated in a much simpler way~\cite{rexloewenDDFT,thisdraft} 
and thus will be conveniently omitted in the following presentation.

Upon substituting the effective force, Eq.~\eqref{eq_effectiveForceEPA}, in the EPA
we rewrite Eq.~\eqref{eq_Smoluchowskieff} as
\begin{align}
\beta\gamma\frac{\partial\PsiF_N}{\partial t}&=\sum_{i=1}^N\nab_i\cdot\sum_{l\neq i}^N
\mathcal{D}^\text{p}(\rr_i-\rr_l)\cdot\bigg(\nab_i\PsiF_N +\PsiF_N\nab_i\beta\Vext^\text{eff}(\bvec{r}_i)
+\PsiF_N\sum_{k\neq i}^N\nab_i \beta u^\text{eff}(\bvec{r}_i,\bvec{r}_k) \bigg). 
\label{eq_appSmoluchowskieffPAIR}
\end{align} 
Integration of this approximate time evolution for the probability density $\PsiF_N(\rr^N,t)$ over $N-1$ coordinates 
yields $\beta\gamma\frac{\partial\rho(\bvec{r},t)}{\partial t}=\nab\cdot\bvec{J}(\bvec{r},t)$ with
the one-body probability current
\begin{align}
\bvec{J}=\!\int\upd\rr''&\,\mathcal{D}^\text{p}(\rr-\rr'')\cdot\bigg(\nab\rho^{(2)}(\rr,\rr'',t)
+\rho^{(2)}(\rr,\rr'',t)\nab\left(\beta\Vext^\text{eff}(\rr)+\beta u^\text{eff}(\rr,\rr'')\right)
+\!\int\upd\rr'\rho^{(3)}(\rr,\rr',\rr'',t)\nab\beta u^\text{eff}(\rr,\rr')\bigg).
\end{align} 
Now we assume that the system evolves in time in such a way that
the time-dependent correlations instantaneously follow those of an equilibrium system.
This adiabatic assumption~\cite{marconi1999,archerevans2004,rexloewenDDFT}
enables us to employ exact equilibrium sum rules
to rewrite the expression in brackets in terms of two-body densities~\cite{rexloewenDDFT,thisdraft}.
The resulting probability current reads
\begin{align}
\bvec{J}=\boldsymbol{\mathcal{D}}(\rr)\cdot\bigg(&\nab\rho(\rr,t)+\rho(\rr,t)\nab\beta\Vext^\text{eff}(\rr)+\int\upd\rr'\rho^{(2)}(\rr,\rr',t)\nab\beta u^\text{eff}(\rr,\rr')\bigg),  \label{eq_appJ}
\end{align} 
where we have defined the ensemble-averaged diffusion tensor
\begin{align}
\boldsymbol{\mathcal{D}}(\rr)=\int\upd\rr''\,\mathcal{D}^\text{p}(\rr-\rr'')\,\frac{\rho^{(2)}(\rr,\rr'',t)}{\rho(\rr,t)}\,,
\end{align}
which we choose to be given by Eq.~\eqref{eq_TeffINV2} of the main text, 
following the result of a calculation in Ref.~\onlinecite{thisdraft} similar to the one presented here.

As a final step, we recognize that the term in brackets in Eq.~\eqref{eq_appJ} 
can be written as $\rho(\bvec{r},t)\delta\beta\mathcal{F}^\text{eff}[\,\rho\,]/\delta\rho(\bvec{r},t)$
in terms of the functional derivative of a free energy functional.
The exact excess free energy (corresponding to the term involving $u^\text{eff}$) 
is only known implicitly and we choose here an approximate representation
with $\mathcal{F}_{\rm ex}^{\rm eff}[\,\rho\,]$ from Eq.~\eqref{meanfield}.
With this identification, the probability current in Eq.~\eqref{eq_appJ} directly leads to
 Eq.~\eqref{eq_DDFT} and Eq.~\eqref{eq_ss3int1} of the main text.

\section{Effective route to the active pressure and interfacial tension\label{app_virial}}

Consider a fluid interacting with the pair potential $u(r)$ confined by an external field 
$\Vext(\bvec{r})$ and define  according to Eq.~\eqref{eq_ss3int1} the average net force as
$\bvec{X}(\bvec{r}):=\boldsymbol{\mathcal{D}}(\bvec{r})\cdot\left(\nab\rho(\bvec{r})+\rho(\bvec{r})\left\langle\nab\beta \mathcal{U}^\text{eff}\right\rangle\right)$ and using the effective potentials defined in appendix~\ref{app_Feff} we obtain:
\begin{align}
\bvec{X}(\bvec{r})=
\boldsymbol{\mathcal{D}}(\bvec{r})
\cdot\left(\nab\rho(\bvec{r})+\rho(\bvec{r})\nab\beta\Vext^\text{eff}(\bvec{r})+
\int\upd\bvec{r}'\rho^{(2)}(\bvec{r},\bvec{r}')\nab\beta u^\text{eff}(|\bvec{r}-\bvec{r}'|)\right)\,,
\label{eq_ss3int1A}
\end{align}
where the diffusion tensor $\boldsymbol{\mathcal{D}}(\bvec{r})$ is given by Eq.~\eqref{eq_TeffINV2}.
Now,  according to Eq.~\eqref{eq_ss3int1}, $\bvec{X}(\bvec{r})\!=\!0$ is a consequence of the steady-state condition,
 Eq.~\eqref{eq_ss1}, so that  Eq.~\eqref{eq_ss3int1A} constitutes an effectively mechanical force-balance condition. 
Our objective is to analyze the equality 
\begin{align}
  \frac{1}{\di V}\int\upd\bvec{r}\bvec{X}(\bvec{r})\cdot\bvec{r}=0 \,,
  \label{eq_Xvirial}
\end{align}
which results from the virial~\eqref{eq_virial} of forces $\bvec{X}(\bvec{r})$.

Recalling that, by the standard external virial, the mechanical pressure is defined as
\begin{align}
 \beta p&=\frac{1}{\di V}\int\upd\bvec{r}\,\rho(\bvec{r}) (\nab\beta\Vext(\bvec{r}))\cdot\bvec{r}\simeq \frac{1}{\di V}\int\upd\bvec{r}\,\rho(\bvec{r})\,\bvec{r}\cdot\boldsymbol{\mathcal{D}}
 (\bvec{r})\cdot\left(\mathcal{D}^{-1}_{[1]}\cdot\nab\beta\Vext(\rr)
 \right)\,.
 \label{eq_Pvirial}
 \end{align}
 We make contact to Eq.~\eqref{eq_Xvirial} in the last step by
inserting the first term of the effective external external potential $\Vext^\text{eff}(\bvec{r})$ from Eq.~\eqref{eq_Veff}.
As discussed in the main text when deriving Eq.~\eqref{eq_YBGid}, this approximate conversion becomes exact
when we consider a non-interacting fluid ($u(r)=u^\text{eff}(r)=0$).
In this case it is easy to show that Eq.~\eqref{eq_Xvirial} yields
\begin{align}\label{eq_PvirialID}
\di V \beta p= -\int\upd\bvec{r}\, (\nab\cdot\rho(\bvec{r})\boldsymbol{\mathcal{D}}(\bvec{r}))\cdot\bvec{r}
=-\int\upd\bvec{r}\, \nab\cdot(\rho(\bvec{r})\boldsymbol{\mathcal{D}}(\bvec{r})\cdot\bvec{r})+\int\upd\bvec{r}\, \rho(\bvec{r})\mbox{Tr}[\boldsymbol{\mathcal{D}}(\bvec{r})]
\end{align}
after integration by parts, 
where the first term on the right-hand side is a boundary term and vanishes.
In the bulk limit, where $\Vext(\bvec{r})\!\rightarrow\!0$ and $\rho(\bvec{r})\!\rightarrow\!\rho$ is constant,
we thus find the active pressure $\beta p=\Da\rho$ of an ideal gas.

Consider now the analogy for the bulk limit of an interacting active fluid.
First we recognize that the second step in Eq.~\eqref{eq_Pvirial} is not exact 
since $\boldsymbol{\mathcal{D}}(\bvec{r})$ depends on the pair potential.
However, it can be easily shown that at linear order in $\tau_\text{a}$, all additional terms depend on both the external field and the pair potential~\cite{thisdraft}.
Secondly, all terms in Eq.~\eqref{eq_ss3int1A} depending on the external field (including these mixed terms) 
will not contribute to the internal bulk stress after taking the limit $\Vext(\bvec{r})\!\rightarrow\!0$
and $\rho(\bvec{r})\!\rightarrow\!\rho$ as for Eq.~\eqref{eq_Pvirial}.
This further means that there exists a wall-independent equation of state,
which is not the case for generic active systems~\cite{solon2015EOS}.
Thus we argue that, in a good approximation, the internal contributions to Eq.~\eqref{eq_Xvirial}
can be calculated with an effective diffusion tensor $\boldsymbol{\mathcal{D}}(\bvec{r})$
whose components 
\begin{align}\label{eq_TeffINV2app}
 \Da \boldsymbol{\mathcal{D}}^{-1}_{\alpha\beta}(\rr)=\delta_{\alpha\beta}+\ttau \int\upd\bvec{r}'\,
\frac{\rho^{(2)}(\bvec{r},\bvec{r}')}{\rho(\bvec{r})}\,\partial_{\alpha}\partial_{\beta}u(r)
\end{align}
 only depend on derivatives $\partial_\alpha=\partial/\partial\alpha$ of the interparticle potential $u(r=|\bvec{r}-\bvec{r}'|)$,
 where the indices $\alpha$ and $\beta$ run over $x$, $y$ and $z$.
 Considering a system that is homogeneous in the $x$-$y$ plane,
 the integrand in Eq.~\eqref{eq_TeffINV2app} is antisymmetric with respect to $\alpha$
 if $\alpha\neq\beta$ (and $\alpha\neq z$) and equally for $\beta$, 
 so that $\boldsymbol{\mathcal{D}}^{-1}$ becomes diagonal and easily invertible.
 The result for $\boldsymbol{\mathcal{D}}_{\alpha\alpha}$ is stated as Eq.~\eqref{eq_virialpressureDii} of the main text.

 Without loss of generality we consider in the following the case of $\di=3$ spatial dimensions.
Separating the result of the external virial from
other contributions, we define for each term of the sum $\sum_\alpha\bvec{X}_\alpha(\bvec{r})\bvec{r}_\alpha$ 
over the vector components in Eq.~\eqref{eq_Xvirial} the diagonal elements 
$\boldsymbol{P}_{\alpha\alpha}(\bvec{r})\!=\!
\boldsymbol{P}^\text{id}_{\alpha\alpha}(\bvec{r})\!+\!\boldsymbol{P}^\text{vir}_{\alpha\alpha}(\bvec{r})$
of the pressure tensor $\boldsymbol{P}(\bvec{r})$, so that $3Vp\!=\!\int\upd\bvec{r}\,\mbox{Tr}[\boldsymbol{P}(\bvec{r})]$.
The first term in Eq.~\eqref{eq_ss3int1A} results in the ideal part
\begin{align}
\beta \boldsymbol{P}^\text{id}_{\alpha\alpha}(\bvec{r})
&=\rho(\bvec{r})\boldsymbol{\mathcal{D}}_{\alpha\alpha}(\bvec{r})+\rho(\bvec{r})\,\bvec{r}_\alpha\partial_\alpha\boldsymbol{\mathcal{D}}_{\alpha\alpha}(\bvec{r})
\label{eq_pressuretensorID}
\end{align}
after performing an integration by parts of
$\int\upd\bvec{r}\beta \boldsymbol{P}^\text{id}_{\alpha\alpha}(\bvec{r})\!:=\!
-\int\upd\bvec{r}\,\bvec{r}_\alpha\boldsymbol{\mathcal{D}}_{\alpha\alpha}(\bvec{r})\partial_\alpha\rho(\bvec{r})$
 as in Eq.~\eqref{eq_PvirialID}
and subsequently taking the bulk limit.
From the third expression in Eq.~\eqref{eq_ss3int1A} we identify
\begin{align}\label{eq_pressuretensorVIR}
 \boldsymbol{P}^\text{vir}_{\alpha\alpha}(\bvec{r})&=-\bvec{r}_\alpha
\boldsymbol{\mathcal{D}}_{\alpha\alpha}(\bvec{r})
\int\upd\bvec{r}'\rho^{(2)}(\bvec{r},\bvec{r}')\partial_\alpha u^\text{eff}(r)=-\bvec{r}_\alpha
\boldsymbol{\mathcal{D}}_{\alpha\alpha}(\bvec{r})
\int\upd\bvec{r}'\rho^{(2)}(\bvec{r},\bvec{r}')\frac{\bvec{r}_\alpha-\bvec{r}_\alpha'}{r}\, \partial_ru^\text{eff}(r)\,,
\end{align}
which is identical to the standard virial contribution for a passive system, 
but here involving the effective interactions (and correlations).

To understand the role of the (uncommon) second term in the ideal pressure tensor~\eqref{eq_pressuretensorID},
we substitute the effective pair potential~\eqref{eq_ueff} into Eq.~\eqref{eq_pressuretensorVIR},
while assuming $\mathcal{D}_{11}(\bvec{r},\bvec{r}')\!\rightarrow\!\boldsymbol{\mathcal{D}}(\bvec{r})$,
 which is only true for single-particle interactions.
 The result is  $\beta\tilde{\boldsymbol{P}}^\text{vir}_{\alpha\alpha}(\bvec{r})\!:=\!
 \beta\boldsymbol{P}^{\text{vir},0}_{\alpha\alpha}(\bvec{r})
 -\rho(\bvec{r})\,\bvec{r}_\alpha\partial_\alpha\boldsymbol{\mathcal{D}}_{\alpha\alpha}(\bvec{r})$,
where the first term 
\begin{align}\label{eq_pressuretensorVIRmarconi0}
 \boldsymbol{P}^{\text{vir},0}_{\alpha\alpha}(\bvec{r}):=
-\bvec{r}_\alpha\int\upd\bvec{r}'\rho^{(2)}(\bvec{r},\bvec{r}')\partial_\alpha u(r)
\end{align}
equals the standard virial formula for a passive system interacting with the bare potential $u(r)$
and the second term is the negative of the expression in $\boldsymbol{P}^\text{id}_{\alpha\alpha}(\bvec{r})$ 
from Eq.~\eqref{eq_pressuretensorID}.
Defining
$\boldsymbol{P}^0\!:=\!
\boldsymbol{P}^\text{id}+\tilde{\boldsymbol{P}}^{\text{vir}}
\!=\!\boldsymbol{P}^{\text{id},0}+\boldsymbol{P}^{\text{vir},0}$
with $\beta\boldsymbol{P}^{\text{id},0}_{\alpha\alpha}(\bvec{r})\!:=\!\rho(\bvec{r})\boldsymbol{\mathcal{D}}_{\alpha\alpha}(\bvec{r})$,
 we recover the YBG results~\cite{marconi2016} for pressure and interfacial tension stated in
Eqs.~\eqref{eq_EFFpYBG} and~\eqref{eq_EFFgammaYBG} of the main text, respectively.
In practice, the contributions to $\boldsymbol{P}$ arising from the effective pair potential
are only partially compensated by the second term in Eq.~\eqref{eq_pressuretensorID},
such that $\boldsymbol{P}$ does not actually reduce to $\boldsymbol{P}^0$.
This cannot be explained alone with the approximations of the EPA (discussed in appendix B of Ref.~\onlinecite{thisdraft}),
as we also ignore higher-order correlations in defining $\boldsymbol{\mathcal{D}}(\bvec{r})$
from a pairwise quantity.
To understand this we notice that we could also recover Eq.~\eqref{eq_pressuretensorVIRmarconi0} from
Eq.~\eqref{eq_pressuretensorVIR} when including $\boldsymbol{\mathcal{D}}(\bvec{r})$ 
to the integrand in the form $\mathcal{D}_{11}(\bvec{r},\bvec{r}')$.
This idea is further elaborated in appendix~\ref{app_pressure2EPA}.
On the other hand, the effective free energy derived from $u^\text{eff}(r)$ in appendix~\ref{app_Feff}
induces higher-order correlations, which are required to accurately describe the effective structure.
Hence, the deviation in $\boldsymbol{P}-\boldsymbol{P}^0$ 
can be interpreted as both an undesired artifact of the EPA and
an approximate compensation accounting for higher-order terms neglected in 
$\boldsymbol{P}^{\text{id},0}$ as we require an approximate pairwise additive diffusion tensor 
to define $\boldsymbol{\mathcal{D}}(\bvec{r})$.

For a homogeneous system, all components $\boldsymbol{\mathcal{D}}_{\alpha\alpha}$ are equal and independent of $\bvec{r}$.
We thus recover Eq.~\eqref{eq_EFFp} of the main text by calculating 
\begin{align}
\beta p=\frac{1}{3}\mbox{Tr}[\boldsymbol{P}]= \frac{1}{3}\mbox{Tr}[\boldsymbol{\mathcal{D}}] (\rho+\beta p^\text{eff}_\text{vir})\simeq \frac{1}{3}\mbox{Tr}[\boldsymbol{\mathcal{D}}] \beta p^\text{eff}
\label{eq_virialpressureA2}
\end{align}
and identifying the effective virial pressure $p^\text{eff}_\text{vir}\simeq p^\text{eff}_\text{ex}$ 
with the excess pressure $p^\text{eff}_\text{ex}$ from DFT, which is equivalent to the compressibility result. 
In practice, this final step is subject to a slight thermodynamic inconsistency, 
which also occurs, e.g., between the solutions of the Percus-Yevick integral equation~\cite{hansen_mcdonald1986}.

Proceeding analogously for the interfacial tension of the planar 
interface,
the diagonal components $\boldsymbol{\mathcal{D}}_{\alpha\alpha}(z)$ in Eq.~\eqref{eq_virialpressureDii} will depend on
the normal coordinate $z$ and thus we have 
$\boldsymbol{\mathcal{D}}_{11}(z)\!=\!\boldsymbol{\mathcal{D}}_{22}(z)\!\neq\!\boldsymbol{\mathcal{D}}_{33}(z)$,
which allows us to differentiate between the tangential $\boldsymbol{P}_\text{T}(z)\!:=\!\boldsymbol{P}_{xx}(z)\!=\!\boldsymbol{P}_{yy}(z)$
and normal $\boldsymbol{P}_\text{N}(z)\!:=\!\boldsymbol{P}_{zz}(z)$ components of the pressure tensor.
The interfacial tension thus becomes
\begin{align}\label{eq_virialinterfaceA1}
\beta \gamma =\frac{1}{A}\int\upd\bvec{r} \,(\boldsymbol{P}_\text{N}(z)-\boldsymbol{P}_\text{T}(z))
&=\!\int\upd z\left( \boldsymbol{\mathcal{D}}_{zz}(z) \,\beta p^\text{eff}_{zz}
+\rho(z)\,z\,\partial_z\boldsymbol{\mathcal{D}}_{zz}(z)
-\boldsymbol{\mathcal{D}}_{xx}(z)\,\beta p^\text{eff}_{xx}(z)\right)\cr&\simeq\!\int\upd z\left(\boldsymbol{\mathcal{D}}_{zz}(z)\,\beta p^\text{eff}
+\boldsymbol{\mathcal{D}}_{xx}(z)\,\omega^\text{eff}(z)+\rho(z)\,z\,\partial_z\boldsymbol{\mathcal{D}}_{zz}(z)\right)
\end{align}
where, according to the definitions in a passive system,
we identify from Eq.~\eqref{eq_pressuretensorVIR} and the first term in Eq.~\eqref{eq_pressuretensorID} the components 
$p^\text{eff}_{\alpha\alpha}\!=\!\rho(z)+\boldsymbol{P}^\text{vir}_{\alpha\alpha}(z)/\boldsymbol{\mathcal{D}}_{\alpha\alpha}(z)$ 
of the effective pressure tensor.
 Note that in a passive system, only the virial stress~\eqref{eq_pressuretensorVIRmarconi0} 
 is relevant to calculate the interfacial tension
 $\beta\gamma_0\!=\!\int\upd z\,(\boldsymbol{P}^{\text{vir},0}_\text{N}(z)-\boldsymbol{P}^{\text{vir},0}_\text{T}(z))$.
  Quite in contrast, the linear term in $\rho(z)$ also contributes to the interfacial tension at non-zero activity,
as $(\boldsymbol{\mathcal{D}}_{zz}(z)-\boldsymbol{\mathcal{D}}_{xx}(z))$ does not vanish.
Translating Eq.~\eqref{eq_virialinterfaceA1} to DFT language in the last step,
similar to Eq.~\eqref{eq_virialpressureA2}, results in Eq.~\eqref{eq_EFFgamma} of the main text.
Here the (constant) effective normal pressure $p^\text{eff}_{zz}\!\simeq\! p^\text{eff}$ corresponds to the bulk pressure
and the effective tangential pressure $p^\text{eff}_{xx}\!\simeq\!-\omega(z)$ to the negative density of the grand potential along the interface.
 Note that $\omega(z)$ explicitly contains the ideal contribution $\rho(z)$, such as all components of the pressure tensor,
whereas the DFT result for the bulk pressure corresponds to the value of the functional for either 
constant density at coexistence and thus cannot be related to the inhomogeneous density profile along the interface.
Therefore, an explicit separation of the interfacial tension into ideal 
and excess contributions as in Eq.~\eqref{eq_virialpressureA2} is not possible.

\section{Alternative rescaling of effective mechanical quantities
\label{app_pressure2EPA}}
In order to derive Eq.~\eqref{eq_DDFT} of the main text, 
we assumed in appendix~\ref{app_DDFT} the effective diffusion tensor to be pairwise additive.
We then argued that, regarding the equivalence of Eqs.~\eqref{eq_ss2} and~\eqref{eq_ss1},
it is convenient to define the averaged diffusion tensor $\boldsymbol{\mathcal{D}}(\bvec{r})$
as the inverse of the average of its inverse.
As a consequence, the formulas we find in appendix~\ref{app_virial}
have the same structure as those derived in Ref.~\onlinecite{marconi2016}
from an expansion of the diffusion tensor $\mathcal{D}_{[N]}(\bvec{r}^N)$ defined in Eq.~\eqref{eq_Gamma0}.
Since we further employ the EPA, the present approach lacks of self-consistency.

 In order to achieve a higher level of self-consistency, we reevaluate 
 the argumentation line from appendix~\ref{app_virial} considering the effective virial pressure tensor
\begin{align}\label{eq_pressuretensorVIR2}
\beta \boldsymbol{P}^{\text{vir}}(\bvec{r})=
-\bvec{r}\cdot
\boldsymbol{\mathcal{D}}
(\bvec{r})\cdot
\int\upd\bvec{r}'\rho^{(2)}(\bvec{r},\bvec{r}') \,\mathcal{D}_{[11]}^{-1}(\bvec{r},\bvec{r}')
\cdot\nab\cdot\left(\boldsymbol{1}\beta u(r)+\mathcal{D}_{[11]}(\bvec{r},\bvec{r}')\right),
\end{align}
 after substituting $u^\text{eff}(r)$ into Eq.~\eqref{eq_pressuretensorVIR}, where
\begin{align}\label{eq_appD11}
\mathcal{D}_{[11]}(\bvec{r},\bvec{r}')=\Da \left(\boldsymbol{1}
+\ttau \Hess u(r)\right)^{-1}\,.
\end{align}
As a limitation of the EPA~\cite{thisdraft}, we recognize artificial three-body correlations in Eq.~\eqref{eq_pressuretensorVIR2}.
At leading order in the activity parameter $\ttau$, however, the term involving
$\nab\cdot\mathcal{D}_{[11]}\propto\ttau$ simplifies dramatically.
This calculation suggests a redefinition of the correction term in 
the ideal pressure tensor from Eq.~\eqref{eq_pressuretensorID} as
\begin{align}
\!\beta \boldsymbol{P}^{\text{id},2}_{\alpha\alpha}(\bvec{r})
=\rho(\bvec{r})\boldsymbol{\mathcal{D}}
_{\alpha\alpha}(\bvec{r})+
\bvec{r}_\alpha\int\upd\bvec{r}'\rho^{(2)}(\bvec{r},\bvec{r}') \,\partial_\alpha
\mathcal{D}_{[11]}(\bvec{r},\bvec{r}')\,.
\label{eq_pressuretensorIDalt}
\end{align}
By doing so, we ensure that at low activity only the first terms in Eqs.~\eqref{eq_pressuretensorVIR2}
and~\eqref{eq_pressuretensorIDalt} contribute to the pressure tensor.
In contrast to $\boldsymbol{P}^{\text{id}}$, this new expression does not vanish in the bulk.

The spurious fact that the two integrations in Eq.~\eqref{eq_pressuretensorVIR2} are carried out independently 
(mind that $\boldsymbol{\mathcal{D}}(\bvec{r})$ is an averaged quantity)
stems from making the EPA~\cite{thisdraft}.
Actually, the integrand of $\boldsymbol{\mathcal{D}}(\bvec{r})$ should cancel with $\mathcal{D}^{-1}_{[11]}$, such that
the term involving $\nab\cdot\boldsymbol{1}\beta u(r)$ reduces to Eq.~\eqref{eq_pressuretensorVIRmarconi0},
which does not depend on the activity at all.
This means that within the EPA we cannot recover the passive virial contribution to the pressure tensor even at leading order in $\ttau$.
However, this consideration suggests that choosing the effective diffusion tensor 
$\boldsymbol{\mathcal{D}}(\bvec{r})\!\approx\!\boldsymbol{\mathcal{D}}^{[2]}(\bvec{r})$ with
 \begin{align}\label{eq_virialpressureDappALT}
\boldsymbol{\mathcal{D}}^{[2]}(\bvec{r})=\int\upd\bvec{r}'\,
\frac{\rho^{(2)}(\bvec{r},\bvec{r}')}{\rho(\bvec{r})}\,\mathcal{D}_{[11]}(\bvec{r},\bvec{r}')
\end{align}
is more consistent with the EPA than the expression in Eq.~\eqref{eq_TeffINV2app}.
 In a manner of speaking we thereby replace the approximate many-body average in Eq.~\eqref{eq_TeffINV2app} 
 with an exact two-body average, Eq.~\eqref{eq_virialpressureDappALT}.

 Again, it is easy to verify that only the diagonal components
    \begin{align}\label{eq_appD11aa}
\left(\mathcal{D}_{[11]}\right)_{\alpha\alpha}(\bvec{r},\bvec{r}')=\Da\,
\frac{1+\ttau \partial_r^2u(r)-
\ttau \!\left(\partial_r^2u(r)-\frac{\partial_ru(r)}{r}\right)\frac{(\bvec{r}_\alpha\!-\bvec{r}_\alpha')^2}{r^2}}
{\left(1+\ttau \partial_r^2u(r)\right)\left(1+\ttau \frac{\partial_ru(r)}{r}\right)}
\,.\ \ \ \ 
\end{align}
of $\boldsymbol{\mathcal{D}}^{[2]}(\bvec{r})$ are important when the system is at most inhomogeneous in the $z$ direction.
 In practice, however, the tensor $\mathcal{D}_{[11]}$ may be ill-defined
 if the validity criteria of the underlying theory are violated~\cite{thisdraft}.
Therefore, a similar correction as for defining the effective potentials would have to be employed in Eq.~\eqref{eq_appD11aa}.

\section{Mean-field free energy in the mechanical picture \label{app_Meanfield}}

In this appendix we derive a steady-state condition of the form \eqref{eq_ss3int1} which gives rise to a different (mean-field) free energy functional. 
Our starting point is the YBG-like expression in Eq.~\eqref{eq_YBGssA1} of the main text, 
which we write here as
\begin{align}
0=\rho(\bvec{r})\langle\nab\beta \mathcal{U}\rangle+\nab\cdot\left(\boldsymbol{\mathcal{D}}(\bvec{r})\rho(\bvec{r})\right)\,,
\label{eq_YBGssA1app}
\end{align}
assuming $\boldsymbol{\mathcal{D}}\simeq\boldsymbol{\mathcal{D}}_\text{I}$ as in \eqref{eq_TeffINV2}. 
This is exact at linear order in $\ttau$~\cite{thisdraft}, i.e., when we define $\boldsymbol{\mathcal{D}}$
from Eq.~\eqref{eq_Gamma0invLT}.
Multiplication of Eq.~\eqref{eq_YBGssA1} with $\boldsymbol{1}\equiv\boldsymbol{\mathcal{D}}\cdot\boldsymbol{\mathcal{D}}^{-1}$ 
does not alter its mechanical character, whereas we may rewrite it in the form
\begin{align}\label{eq_YBGssA2}
0&=\boldsymbol{\mathcal{D}}(\bvec{r})\cdot\rho(\bvec{r})\left(\nab\ln\rho(\bvec{r})+\boldsymbol{\mathcal{D}}^{-1}(\bvec{r})\cdot\langle\nab\beta \mathcal{U}\rangle-\nab\ln(\det\boldsymbol{\mathcal{D}}^{-1}(\bvec{r}))\right)\cr
&\approx\boldsymbol{\mathcal{D}}(\bvec{r})\cdot\left(\rho(\bvec{r})\nab\left(
\frac{\delta\beta\mathcal{F}_\text{id}[\,\rho\,]}{\delta\rho(\bvec{r})}+
\frac{\delta\beta\tilde{\mathcal{F}}^{(1)}_\text{ex}[\,\rho\,]}{\delta\rho(\bvec{r})}+
\frac{\delta\beta\tilde{\mathcal{F}}^{(2)}_\text{ex}[\,\rho\,]}{\delta\rho(\bvec{r})}\right)\right)
\end{align}
 resembling Eq.~\eqref{eq_ss3int1} of the main text.
In the last step we identify a free energy functional
$\tilde{\mathcal{F}}\!=\!\mathcal{F}_\text{id}+\tilde{\mathcal{F}}^{(1)}_\text{ex}+\tilde{\mathcal{F}}^{(2)}_\text{ex}$,
such that the left-hand-side arises from its functional derivative.

The first term corresponds to the standard ideal-gas free energy
$\tilde{\mathcal{F}}_\text{id}[\,\rho\,]\!=\!\int \mathrm{d}\bvec{r}\, \rho(\bvec{r})\left(\,\ln(\Lambda^3\rho(\bvec{r}))-1\right)$.
The second and third terms contain the inverse of the ensemble-averaged diffusion tensor $\boldsymbol{\mathcal{D}}^{-1}$, 
which for an interacting system depends on the pair density.
To obtain  the free energy as a functional of the one-body density alone,
we employ the simplest mean-field approximation $\rho^{(2)}(\bvec{r},\bvec{r}')\approx\rho(\bvec{r})\rho(\bvec{r}')$ 
in \eqref{eq_TeffINV2} and identify
\begin{align}
\frac{\delta\beta\tilde{\mathcal{F}}^{(1)}_\text{ex}[\,\rho\,]
}{\delta\rho(\bvec{r})}&= \frac{\beta}{\Da}\left(\bar{U}(\bvec{r})
+\frac{\ttau}{2}\left(\nab\bar{U}(\bvec{r})\right)^2\right),\label{eq_appF1}\\
\frac{\delta\beta\tilde{\mathcal{F}}^{(2)}_\text{ex}[\,\rho\,]
}{\delta\rho(\bvec{r})}&=- 
\ln\det\left(\boldsymbol{1}+\ttau \Hess\beta\bar{U}(\bvec{r})\right),
\label{eq_appF2}
\end{align}
where we have defined $\bar{U}(\bvec{r})=\Vext(\bvec{r})+\!\int \mathrm{d}\bvec{r}'\, \rho(\bvec{r}')u(\bvec{r},\bvec{r}')$.
By construction of the free energy $\tilde{\mathcal{F}}$ we recover the form of the mechanical condition in Eq.~\eqref{eq_YBGssA1app},
but at the cost of neglecting all correlations.
For a non-interacting system a functional integration of Eqs.~\eqref{eq_appF1} and~\eqref{eq_appF2} becomes possible
and $\tilde{\mathcal{F}}\!\equiv\!\mathcal{F}$ becomes exact and thus equivalent to the free energy derived in appendix~\ref{app_Feff} within the EPA.

There is no analytic expression for $\tilde{\mathcal{F}}$ in the general interacting case, 
which means that the EPA is more useful.
Moreover, with the notion of an effective potential, it is easy to construct a closed theory beyond the simplest mean-field level,
e.g., by treating the repulsive part employing an elaborate functional for hard spheres~\cite{wittmannbrader2016,rosenfeld89}.
 In contrast, the structure predicted from the free energy $\tilde{\mathcal{F}}$
is highly inconsistent wih the mean-field assumption $g(r)\!\equiv\!1$.
However, in the presence of both an external field and interparticle interactions, Eqs.~\eqref{eq_appF1} and~\eqref{eq_appF2}
account for the a coupling between the corresponding potentials, which is ignored in the EPA.
Connecting to the discussion in appendix~\ref{app_pressure2EPA}, we recognize that at linear order in $\ttau$
the expression in Eq.~\eqref{eq_appF2} is also found in the EPA when we additionally make the same mean-field approximation.

\end{document}